# Perturbed locations and stability of the Triangular equilibrium points in the multi-dissipative photogravitational triaxial elliptic RTBP

F. A. Abd El-Salam[1,2]

[1] Department of Mathematics, Faculty of Science, Taibah University, Madina, KSA.

[2] Department of Astronomy, Faculty of Science, Cairo University, Cairo,12613, Egypt.

E.Mail address  f.a.abdelsalam@gmail.com

**Abstract**

The multi-dissipative photogravitational triaxial elliptic restricted three body problem is treated. The perturbed locations of triangular points are computed. The stability of the triangular points under changing one or more perturbing parameter is investigated. The results revealed that at certain values of the considered perturbing parameters, we haven't triangular equilibrium points, or at least there exists but very far from the origin. The change in the stability/instability regions with eccentricity seems to be nonlinear. It seems that increasing the eccentricity the enlarge the stability regions and vice versa. It is revealed that we have two disjoint stability regions. The stability/instability regions for a new set of eccentricities are merged to one region. Consider a very high eccentricity stability region for the whole domain of the mass ratio except in the neighborhood of 0.3 is obtained. stability/asymptotic stability/instability regions due to certain photogravitational effects correspond are revealed.

## 1. Introduction

The restricted three body problem (in brief RTBP) describes the motion of an infinitesimal mass $m_3$ moving under the gravitational effect of the two massive primaries of masses $m_1$ & $m_2$ such that $m_2 \ll m_1$. These primaries are assumed move in circular orbits around their centre of mass on account of their mutual attraction and the infinitesimal mass not influencing the motion of the primaries. If the primaries are assumed move in elliptic orbits, thus we have the so called elliptic restricted three body problem (in brief ERTBP). The ERTBP is of much greater complexity. The main reasons are, on the one hand the generalized motion of the primaries and on the other hand, the fact, that the Hamiltonian becomes time-dependent. The ERTBP does not possess Jacobi integral. But it can be reduced to a mathematically similar form as the circular problem using the non-uniformly rotating and pulsating coordinate reference frame. This pulsating system can be introduced by scaling the distances with respect to the variable mutual distance between the primaries as a unit of length. The system can again be brought into a form where the positions of the primaries are fixed and the motion of the third body can be analyzed relative to the fixed primary locations. However contrary to the circular case, the equations of motion are no longer autonomous since the Hamiltonian still depends on the new independent variable, namely the true anomaly $f$.

Abd El-Salam (2015) treated the elliptic restricted three-body problem with oblate and triaxial primaries. He obtained a new expressions for the triangular locations as power series in the mass ratio parameter. He investigated the linear stability of the triangular points. To avoid the repetition and lengthy introduction, the reader is advised to refer to this article to see the literature is wealth with works dealing with the circular as well as elliptic restricted three body problem with or without some considered perturbations.

Now, it seems worth to sketch some of the most important works focus on the effects of the dissipative force in the restricted three body problem. Several eminent authors but not conclusive, e.g. Colombo et al. (1966), Chernikov (1970), Schuerman (1980) and Liou. et al. (1995) discussed the position as well as the stability of the Lagrangian equilibrium points using Poynting Robertson drag. Murray (1994) systematically discussed the dynamical effect of general drag in the planar circular restricted three body problem. Jain, et al. (2014) performed an analysis of the dynamics of the circular restricted three body problems under the effect of the dissipative force and Poynting Robertson drag. They investigated the existence and stability of stationary solutions. Umar and Singh (2014) investigated the effects of oblateness, radiation and eccentricity of both primaries on the periodic orbits around the triangular Lagrangian points of oblate and luminous binary systems in the framework of the ERTBP. Narayan and Shrivastava (2013,) discussed the oblateness and the photogravitational effects of both the primaries on the location and the stability of the triangular equilibrium points in the ERTBP. They studied the stability of the triangular points under the photogravitational and oblateness effects of both the primaries around the binary systems Achird, Lyeten, Alpha Cen-AB, Kruger 60, and Xi-Bootis.

The article map is as follows: In section 2, the equations of motion of an infinitesimal mass in the multi-dissipative photogravitational triaxial ERTBP are presented. In section 3 and its sebsequent subsections, we computed the locations of triangular equilibrium points, and analyzed the perturbed locations under some considered perturbations. In section 4 we investigated the linear stability of the triangular points and represented the stability/instability regions for some different cases. In section 5, we concluded the obtained results.

## 2. Multi-dissipative photogravitational triaxial ERTBP

The equations of motion of an infinitesimal mass in the elliptic restricted three body problem with oblate and triaixal primaries in a dimensionless, barycentric, pulsating rotating, coordinate system are given by

$$\frac{d^2\xi}{df^2} - 2\frac{d\eta}{df} = \mathcal{E}\frac{\partial U}{\partial \xi} + F_\xi^{SN} + F_\xi^{PR} + F_\xi^{NG}, \qquad (1)$$

$$\frac{d^2\eta}{df^2} + 2\frac{d\xi}{df} = \mathcal{E}\frac{\partial U}{\partial \eta} + F_\eta^{SN} + F_\eta^{PR} + F_\eta^{NG} \qquad (2)$$

where upon averaging $\mathcal{E} = (1-e^2)^{-1/2}$ and

$$U = \frac{1}{2}\left[q(1-\mu)r_1^2 + \mu r_2^2\right] + \frac{1}{n^2}\left[\frac{q(1-\mu)}{r_1} + \frac{A_1 q(1-\mu)}{2r_1^3}\right.$$
$$\left. + \frac{q(1-\mu)(2\sigma_1 - \sigma_2)}{2r_1^3} + \frac{\mu}{r_2} + \frac{A_2 \mu}{2r_2^3} + \frac{\mu(2\gamma_1 - \gamma_2)}{2r_2^3}\right] \quad (3)$$

where $q_i$ ( $i = 1, 2$) denote the respective radiation factors for the bigger and smaller primaries. the numerical values of these parameters are $0 < 1 - q_i \ll 1$, ($i = 1, 2$). $q_i = 1 - (F_p / F_g)_i$, ( $i = 1, 2$) is the mass reduction factor constant for a given mass. $F_p$, being the force due to radiation pressure, $F_g$ being the force due to gravitational field. As the solar radiation pressure force $F_p$ is exactly opposite to the gravitational attraction force $F_g$ ( neglecting a tiny effect of aberration ) and changes with the distance by the same law, it is possible to consider that the result of action of this force will lead to reducing the effective mass of the primaries.

The force due multi-dissipations are $F_\sigma^{SN}, F_\sigma^{PR}, F_\sigma^{NG}, \sigma \equiv \xi, \eta$, they represent the components of the simple nebular drag, Poynting-Robertson drag (in brief PR) and nebular gas (Stokes) drag respectively. They are given by Dvorak and Lhotka (2013) as,

$$F_\xi^{SN} = k_{SN} \dot{\xi} r^\gamma \quad (4)$$

$$F_\eta^{SN} = k_{SN} \dot{\eta} r^\gamma \quad (5)$$

$$F_\xi^{PR} = \frac{k_{PR}}{r_1^2}\left[\dot{\xi} - \eta + \frac{\xi}{r_1^2}(\xi\dot{\xi} + \eta\dot{\eta})\right] \quad (6)$$

$$F_\eta^{PR} = \frac{k_{SN}}{r_1^2}\left[\dot{\eta} + \xi + \frac{\eta}{r_1^2}(\xi\dot{\xi} + \eta\dot{\eta})\right] \quad (7)$$

$$F_\xi^{NG} = -k_{NG}\rho\sqrt{(\dot{\xi} + (\Omega_g - 1)\eta)^2 + (\dot{\eta} + (1-\Omega_g)\xi)^2} \ (\dot{\xi} - (1-\Omega_g)\eta) \quad (8)$$

$$F_\eta^{NG} = k_{NG}\rho\sqrt{(\dot{\xi} - (1-\Omega_g)\eta)^2 + (\dot{\eta} + (1-\Omega_g)\xi)^2} \ (\dot{\eta} + (1-\Omega_g)\xi) \quad (9)$$

and $k's$, $\gamma$ are the model parameters. The density $\rho = \rho(r)$ and angular velocity of the gas $\Omega_g = \Omega_g(r)$ are functions depending on the radial component $r$ also.

The PR drag, $F_\sigma^{PR}$, $\sigma \equiv \xi, \eta$, is a force that is proportional to the inverse square distance $1/r_1^2$ from the radiant massive primary (e.g. Sun). The first two terms inside the brackets are due to the impact of the photons of the solar radiation with the particle. They are proportional to the velocity of the infinitesimal mass in the synodic frame. The coefficient of the PR drag effect is of the order of $10^{-5}$. Due to the fact the particle is moving around the Sun, the last term, that proportional to $1/r_1^4$, represents the Doppler shift of the solar radiation that hits the particle, see Jain, et al. (2014).

Finally $a$ and $e$ are the semi major axis and eccentricity of either primary respectively and $f$ is the true anomaly of the $m_1$, $(A_i, \sigma_i, \gamma_i \ll 1)$, $i=1,2$ are the oblateness and the triaxial coefficients of the bigger and smaller primaries respectively. Assume that the principal axes of $m_1$ and $m_2$ are parallel to the synodic axes defined by Szebehely (1967) and $a_{m_1}, b_{m_1}, c_{m_1}, a_{m_2}, b_{m_2}, c_{m_2}$ are the lengths of the semi-axes of bigger and smaller primaries respectively, then

$$\sigma_1 = \frac{a_{m_1}^2 - c_{m_1}^2}{5r^2}, \sigma_1 = \frac{b_{m_1}^2 - c_{m_1}^2}{5r^2}, \gamma_1 = \frac{a_{m_2}^2 - c_{m_2}^2}{5r^2}, \gamma_2 = \frac{b_{m_2}^2 - c_{m_2}^2}{5r^2}.$$

Also $r_1 = \sqrt{(\xi+\mu)^2 + \eta^2}$ and $r_2 = \sqrt{(\xi+\mu-1)^2 + \eta^2}$ are the distances of the infinitesimal mass from these primaries in the rotating-pulsating coordinates, $\mu$ is the ratio of the mass of the smaller primary to the total mass of the primaries and $0 < \mu \leq 1/2$, and $n$, the mean motion, is given by Abd El-Salam (2015)

$$n^2 = \frac{1}{a(1-e^2)}\left[1 + \frac{3qA_1}{2} + \frac{3A_2}{2} + \frac{3}{2}q(2\sigma_1 - \sigma_2) + \frac{3}{2}(2\gamma_1 - \gamma_2)\right] \qquad (10)$$

### 3. Locations of triangular equilibrium points

The positions of the equilibrium points can be found by setting all relative velocity and relative acceleration components equal to zero and solving the resulting system,

$$\left[\mathcal{E}U_\xi + F_\xi^{SN} + F_\xi^{PR} + F_\xi^{NG}\right]_{\dot{\xi}=\dot{\eta}=0} = \left[\mathcal{E}U_\eta + F_\eta^{SN} + F_\eta^{PR} + F_\eta^{NG}\right]_{\dot{\xi}=\dot{\eta}=0} = 0 \qquad (11)$$

Where

$$\mathcal{E}\left[n^2\xi - \frac{q(1-\mu)(\xi+\mu)}{r_1^3} - \frac{3A_1 q(1-\mu)(\xi+\mu)}{2r_1^5} - \frac{3q(1-\mu)(2\sigma_1-\sigma_2)(\xi+\mu)}{2r_1^5}\right.$$

$$\left. - \frac{\mu(\xi+\mu-1)}{r_2^3} - \frac{3A_2\mu(\xi+\mu-1)}{2r_2^5} - \frac{3\mu(2\gamma_1-\gamma_2)(\xi+\mu-1)}{2r_2^5}\right]$$

$$-k_{PR}\frac{\eta}{r_1^2} - k_{NG}k\left(1-\Omega_g\right)^2 \rho r\eta = 0$$

and

$$\varepsilon n^2\eta - \varepsilon\eta\left\{\frac{q(1-\mu)}{r_1^3} + \frac{3q(1-\mu)A_1}{2r_1^5} + \frac{3q(1-\mu)(2\sigma_1-\sigma_2)}{2r_1^5} + \frac{\mu}{r_2^3} - \frac{3\mu A_2}{2r_2^5} + \frac{3\mu(2\gamma_1-\gamma_2)}{2r_2^5}\right\}$$

$$+k_{PR}\frac{\xi}{r_1^2} + k_{NG}\left(1-\Omega_g\right)^2 \rho r\xi = 0$$

Setting $A_\sigma = \left(A_1 + 2\sigma_1 - \sigma_2\right)$, $A_\gamma = \left(A_2 + 2\gamma_1 - \gamma_2\right)$, then the above two equations can be written concisely as

$$\varepsilon\left[n^2\xi - \frac{q(1-\mu)(\xi+\mu)}{r_1^3} - \frac{3qA_\sigma(1-\mu)(\xi+\mu)}{2r_1^5} - \frac{\mu(\xi+\mu-1)}{r_2^3} - \frac{3A_\gamma\mu(\xi+\mu-1)}{2r_2^5}\right]$$

$$-k_{PR}\frac{\eta}{r_1^2} - k_{NG}\left(1-\Omega_g\right)^2 \rho r\eta = 0 \qquad (12)$$

and

$$\varepsilon n^2\eta - \varepsilon\eta\left\{\frac{q(1-\mu)}{r_1^3} + \frac{3qA_\sigma(1-\mu)}{2r_1^5} + \frac{\mu}{r_2^3} + \frac{3\mu A_\gamma}{2r_2^5}\right\} + k_{PR}\frac{\xi}{r_1^2} + k_{NG}\left(1-\Omega_g\right)^2 \rho r\xi = 0 \qquad (13)$$

Since the oblateness and triaxiality coefficients are small, i.e. $A_\sigma, A_\gamma \ll 1$, $i=1,2$. Ignoring these perturbations yields the equilateral solution of the classical restricted three-body problem i.e. $r_1 = r_2 = 1$, then it may be reasonable in our case to assume that the positions of the equilibrium points $L_{4,5}$ are the same as given by classical restricted three-body problem but perturbed by terms factored by $\varepsilon_{1,2} \equiv \mathcal{O}(A_\sigma, A_\gamma)$. i.e $r_i = 1 + \varepsilon_i$, $i=1,2$. Substituting this assumption in $r_1 = \sqrt{(\xi+\mu)^2+\eta^2}$, and $r_2 = \sqrt{(\xi+\mu-1)^2+\eta^2}$ then solving for $\xi$ and $\eta$ up to the first order in the involved small quantities $\varepsilon_1$ and $\varepsilon_2$ we get

$$\xi = \varepsilon_1 - \varepsilon_2 + (1-2\mu)/2, \qquad \eta = \pm\left[(\sqrt{3}/2) + (\sqrt{3}/3)(\varepsilon_1+\varepsilon_2)\right] \qquad (14)$$

Substituting the values of $r_1$, $r_2$, $\xi$ and $\eta$ into equations (12) and (13), and expanding the resulting equations, we can ignore all the higher order terms in $\varepsilon_1$ and $\varepsilon_2$ than the first order, as well as the mixed terms. Thus we get the following two simultaneous equations in $\varepsilon_1$, $\varepsilon_2$.

$$\mathcal{A}_1 \varepsilon_1 + \mathcal{B}_1 \varepsilon_2 + S_1 = 0 \;, \qquad\qquad \mathcal{A}_2 \varepsilon_1 + \mathcal{B}_2 \varepsilon_2 + S_2 = 0$$

where

$$\begin{aligned}
\mathcal{A}_1 =\ & n^2 + \frac{q}{2} + \frac{9}{4} q A_\sigma - \rho k_{NG} + 2\rho k_{NG} \Omega_g - \rho k_{NG} \Omega_g^2 \\
& - \mu - \frac{1}{2} q\mu - \frac{21}{4} A_\gamma \mu - \frac{9}{4} q A_\sigma \mu + \frac{3}{2} \rho k_{NG} \mu - 2 k_{PR} \mu - 3\rho k_{NG} \Omega_g \mu \\
& + \frac{3}{2} \rho k_{NG} \Omega_g^2 \mu - 2\rho k_{NG} \mu^2 + 4\rho k_{NG} \Omega_g \mu^2 - 2\rho k_{NG} \Omega_g^2 \mu^2
\end{aligned}$$

$$\begin{aligned}
\mathcal{B}_1 =\ & -n^2 + q + \frac{3}{2} q A_\sigma + \rho k_{NG} + k_{PR} - 2\rho k_{NG} \Omega_g + \rho k_{NG} \Omega_g^2 \\
& - \frac{\mu}{2} - q\mu + \frac{3}{2} A_\gamma \mu - \frac{3}{2} q A_\sigma \mu - \frac{3}{2} \rho k_{NG} \mu + 3\rho k_{NG} \Omega_g \mu \\
& - \frac{3}{2} \rho k_{NG} \Omega_g^2 \mu + 2\rho k_{NG} \mu^2 - 4\rho k_{NG} \Omega_g \mu^2 + 2\rho k_{NG} \Omega_g^2 \mu^2
\end{aligned}$$

$$\begin{aligned}
S_1 =\ & \frac{1}{2} n^2 - \frac{1}{2} q - \frac{3}{4} q A_\sigma - \frac{1}{2} k_{PR} - \frac{1}{2} \rho k_{NG} + \rho k_{NG} \Omega - \frac{1}{2} \rho k_{NG} \Omega_g^2 \\
& - \frac{1}{2} n^2 \mu + \frac{1}{2} q\mu + \frac{3}{4} q A_\sigma \mu + \frac{1}{2} \mu + \frac{3}{4} A_\gamma \mu + k_{PR} \mu + \frac{3}{2} \rho k_{NG} \mu \\
& - 3\rho k_{NG} \Omega_g \mu + \frac{3}{2} \rho k_{NG} \Omega_g^2 \mu - \frac{3}{2} \rho k_{NG} \mu^2 + 3\rho k_{NG} \Omega_g \mu^2 \\
& - \frac{3}{2} \rho k_{NG} \Omega_g^2 \mu^2 + \rho k_{NG} \mu^3{}_g - 2\rho k_{NG} \Omega_g \mu^3 + \rho k_{NG} \Omega_g^2 \mu^3
\end{aligned}$$

$$\begin{aligned}
\mathcal{A}_2 =\ & -\frac{n^2}{\sqrt{3}} + \frac{11 q}{2\sqrt{3}} + \rho k_{NG} + \frac{17}{4} \sqrt{3} q A_\sigma - 2\rho k_{NG} \Omega_g + \rho k_{NG} \Omega_g^2 \\
& + \frac{1}{\sqrt{3}} \mu - \frac{11}{2\sqrt{3}} q\mu - \frac{17}{4} \sqrt{3} A_\gamma \mu - \frac{17}{4} \sqrt{3} q A_\sigma \mu \\
& - \frac{3}{2} \rho k_{NG} \mu + 2 k_{PR} \mu + 3\rho k_{NG} \Omega_g \mu - \frac{3}{2} \rho k_{NG} \Omega_g^2 \mu \\
& + 2\rho k_{NG} \mu^2 - 4\rho k_{NG} \Omega_g \mu^2 + 2\rho k_{NG} \Omega_g^2 \mu^2
\end{aligned}$$

$$\begin{aligned}
\mathcal{B}_2 =\ & \frac{1}{\sqrt{3}} n^2 - \frac{1}{\sqrt{3}} q - \frac{1}{2} \sqrt{3} q A_\sigma - \rho k_{NG} - k_{PR} + 2\rho k_{NG} \Omega_g - \rho k_{NG} \Omega_g^2 \\
& + \frac{7}{2\sqrt{3}} \mu + \frac{q}{\sqrt{3}} \mu + \frac{1}{2} \sqrt{3} A_\gamma \mu + \frac{1}{2} \sqrt{3} q A_\sigma \mu + \frac{3}{2} \rho k_{NG} \mu - 3\rho k_{NG} \Omega_g \mu \\
& + \frac{3}{2} \rho k_{NG} \Omega_g^2 \mu - 2\rho k_{NG} \mu^2 + 4\rho k_{NG} \Omega_g \mu^2 - 2\rho k_{NG} \Omega_g^2 \mu^2
\end{aligned}$$

$$S_2 = \frac{\sqrt{3}}{2}n^2 - \frac{\sqrt{3}}{2}q - \frac{3\sqrt{3}}{4}qA_\sigma + \frac{\rho k_{NG}}{2} + \frac{k_{PR}}{2} - \rho k_{NG}\Omega_g + \frac{1}{2}\rho k_{NG}\Omega_g^2$$
$$- \frac{\sqrt{3}}{2}\mu + \frac{\sqrt{3}}{2}q\mu + \frac{3\sqrt{3}}{4}A_\gamma\mu + \frac{3\sqrt{3}}{4}qA_\sigma\mu - \frac{3}{2}\rho k_{NG}\mu - \frac{3}{2}\rho k_{NG}\Omega_g^2\mu$$
$$- k_{PR}\mu + 3\rho k_{NG}\Omega_g\mu - 3\rho k_{NG}\Omega_g\mu^2 + \frac{3}{2}\rho k_{NG}\mu^2 + \frac{3}{2}\rho k_{NG}\Omega_g^2\mu^2$$
$$- \rho k_{NG}\mu^3 + 2\rho k_{NG}\Omega_g\mu^3 - \rho k_{NG}\Omega_g^2\mu^3$$

which represent two simultaneous equations in $\varepsilon_1$, $\varepsilon_2$ their solution gives

$$\varepsilon_1 = -\frac{S_1\mathcal{B}_2 - S_2\mathcal{B}_1}{\mathcal{A}_1\mathcal{B}_2 - \mathcal{A}_2\mathcal{B}_1}, \qquad \mathcal{A}_1\mathcal{B}_2 - \mathcal{A}_2\mathcal{B}_1 \neq 0$$

$$\varepsilon_2 = \frac{\mathcal{A}_2 S_1 - \mathcal{A}_1 S_2}{\mathcal{A}_1\mathcal{B}_2 - \mathcal{A}_2\mathcal{B}_1}, \qquad \mathcal{A}_1\mathcal{B}_2 - \mathcal{A}_2\mathcal{B}_1 \neq 0$$

Which can be written as a function up to order five in $\mu$ as

$$\varepsilon_1 = -\frac{1}{\mathcal{R}}\sum_{k=-1}^{5}\mathcal{D}_{1,k}\mu^k, \quad \varepsilon_2 = \frac{1}{\mathcal{R}}\sum_{k=-1}^{5}\mathcal{D}_{2,k}\mu^k, \quad \forall\, \mathcal{R} = \sum_{k=-1}^{5}\mathcal{D}_{0,k}\mu^k \neq 0$$

Where the non-vanishing coefficients $\mathcal{D}_{0,k}$, $\mathcal{D}_{1,k}$ and $\mathcal{D}_{2,k}$ are given in appendix A.

Substituting the values of $\varepsilon_1$, $\varepsilon_2$ into equation (14) yields the coordinates of the triangular points

$$\xi_{L_{4,5}} = \frac{1}{2} - \mu - \frac{1}{\mathcal{R}}\sum_{k=-1}^{5}\left(\mathcal{D}_{1,k} + \mathcal{D}_{2,k}\right)\mu^k \tag{15}$$

$$\eta_{L_{4,5}} = \pm\left[\frac{\sqrt{3}}{2} - \frac{\sqrt{3}}{3}\frac{1}{\mathcal{R}}\sum_{k=-1}^{5}\left(\mathcal{D}_{1,k} - \mathcal{D}_{2,k}\right)\mu^k\right] \tag{16}$$

### 3.1. Graphical representations

In Fig. 1, 3, 5 we plotted the location of $L_{4,5}$ points in ERTBP without inclusion the oblateness and triaxial effects. We included in all figures a constant PR drag. In Fig. 1 we assumed that the primaries do not radiate. The black curves represent the nebular gas drag free problem but other curves blue, red and green represent increasing drag values for increasing eccentricities of the primaries orbits. As is clear from the figure, the change is small for and it may be considered as linear decreasing variation for $\mu \in [0, 0.3)$. This situation converses for $\mu > 0.3$ for large drag values at a relatively moderate eccentricities, see the green curves. In Fig. 2 we treated the same case as in Fig. 1 but with inclusion the oblateness and triaxial effects. The remarkable feature is that: at certain values of the considered variable we haven't triangular equilibrium points, as some curves are not continuous, see specially all green curves and some red ones. In Fig. 3 we plotted the location of $L_{4,5}$ points in circular RTBP. The black curves represent the nebular gas drag free problem but other curves blue, red and green represent increasing drag values for increasing photogravitational effects of the primaries. As is clear from the figure, the change is small and it may be considered as qusi-linear decreasing variation for the very low drag values, see the black and the blue curves. While for large drag values and a relatively high radiation, see the red and green curves, the effect becomes nonlinear and the location of $L_{4,5}$ increase respectively. In Fig. 4 we treated the same case as in Fig. 3 but with inclusion the oblateness and triaxial effects. The dynamics are nearly the same but with different sizes of resulting perturbations. In Fig. 5 we plotted the location of $L_{4,5}$ points in ERTBP. The black curves represent the non radiant ERTBP but other curves blue, red and green represent photograitational ERTBP with increasing of mass transfer by radiation for increasing eccentricities of the primaries orbits. As is clear from the figure, the change is small and it may be considered as linear decreasing variation for non radiant circular RTBP. But for a relatively high radiation and eccentricity, see the red and green curves, the effect becomes nonlinear and the location of $L_{4,5}$ is rapidly increasing respectively. In Fig. 6 we treated the same case as in Fig. 5 but with inclusion the oblateness and triaxial effects. The dynamics are nearly the same but with different sizes of resulting perturbations.

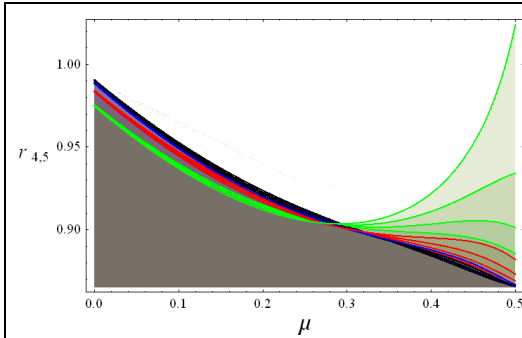

Fig. 1: The Locations of $L_{4,5}$ with

$\rho \in [0.0, 0.3], e \in [0.0, 0.3], q = 1.0$
$k_{NG} = 0.0, \Omega_g = 0.0, k_{PR} = 0.5$

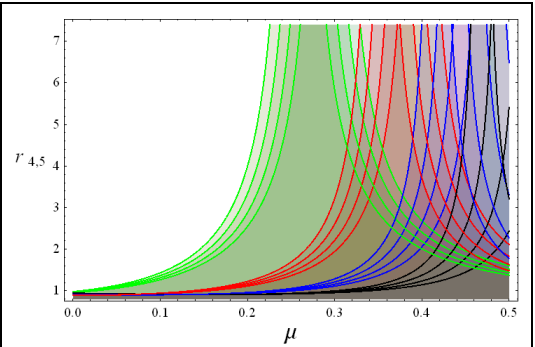

Fig. 2: The Locations of $L_{4,5}$ with

$\rho \in [0.0, 0.3], e \in [0.0, 0.3], q = 1.0$
$k_{NG} = -0.5, \Omega_g = 1 \times 10^{-4}, k_{PR} = 0.5$

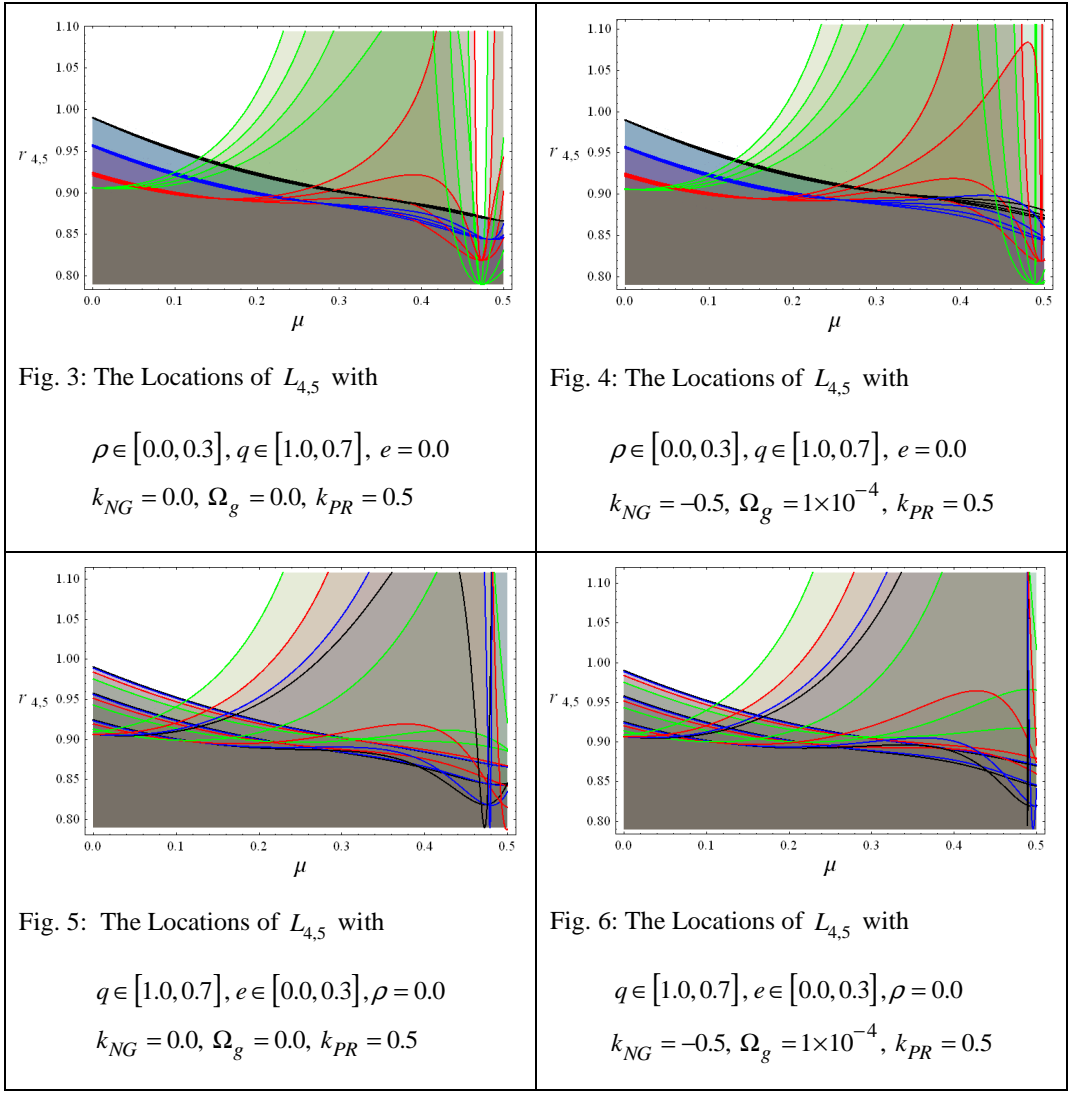

Fig. 3: The Locations of $L_{4,5}$ with

$\rho \in [0.0, 0.3], q \in [1.0, 0.7], e = 0.0$
$k_{NG} = 0.0, \Omega_g = 0.0, k_{PR} = 0.5$

Fig. 4: The Locations of $L_{4,5}$ with

$\rho \in [0.0, 0.3], q \in [1.0, 0.7], e = 0.0$
$k_{NG} = -0.5, \Omega_g = 1 \times 10^{-4}, k_{PR} = 0.5$

Fig. 5: The Locations of $L_{4,5}$ with

$q \in [1.0, 0.7], e \in [0.0, 0.3], \rho = 0.0$
$k_{NG} = 0.0, \Omega_g = 0.0, k_{PR} = 0.5$

Fig. 6: The Locations of $L_{4,5}$ with

$q \in [1.0, 0.7], e \in [0.0, 0.3], \rho = 0.0$
$k_{NG} = -0.5, \Omega_g = 1 \times 10^{-4}, k_{PR} = 0.5$

### 3.2. Analysis of photogravitational effects on $L_{4,5}$

On Figs. 7-11, we plotted the location of $L_{4,5}$ points in ERTBP, assuming constant nebular and PR drag. we applied our case study for constant eccentricity of primaries orbit $e = 0.2$. We considered a wide range of photogravitational perturbing parameter $q$ for different mass ratios. The curve colors: blue, red. purple, cyan, green indicate the cases from $q = 0.9$ to $q = 0.5$ respectively. Fig. 7. shows the photogravitational effect on the triangular points $L_{4,5}$ for systems with $\mu = 0.1$. As is clear from the figure, the bigger the mass loss the larger the shift in $L_{4,5}$ location. It seems that the case with $q = 0.5$, $\mu = 0.1$ has no triangular equilibrium point, or at least there exists but very far from the barycenter, see the discontinuity of the green curve. The reason that enhance this interpretation is the reflection of the green curve to the right in Fig. 8. In Fig. 9, $\mu = 0.3$ the cases $q = 0.5, q = 0.6, q = 0.7$ we cannot properly conclude the existence of the triangular equilibrium points. In Fig. 10, $\mu = 0.4$ we have a family of triangular equilibrium points, four of them exist to the left of the $\eta$-axis in the second and third quadrants. These points correspond to the cases $q = 0.9, q = 0.8$. Other four equilibrium points correspond to $q = 0.6, q = 0.5$ exist to the right of the $\eta$-axis in the first and fourth quadrants. This may be attributed to the interchange of the locations of the more and less massive primaries due to the severe radiation cases. Fig. 11 shows the locations of $L_{4,5}$ for the a systems of equal masses with different photogravitational parameters. All five cases are represented in the rectangular Fig. 12.

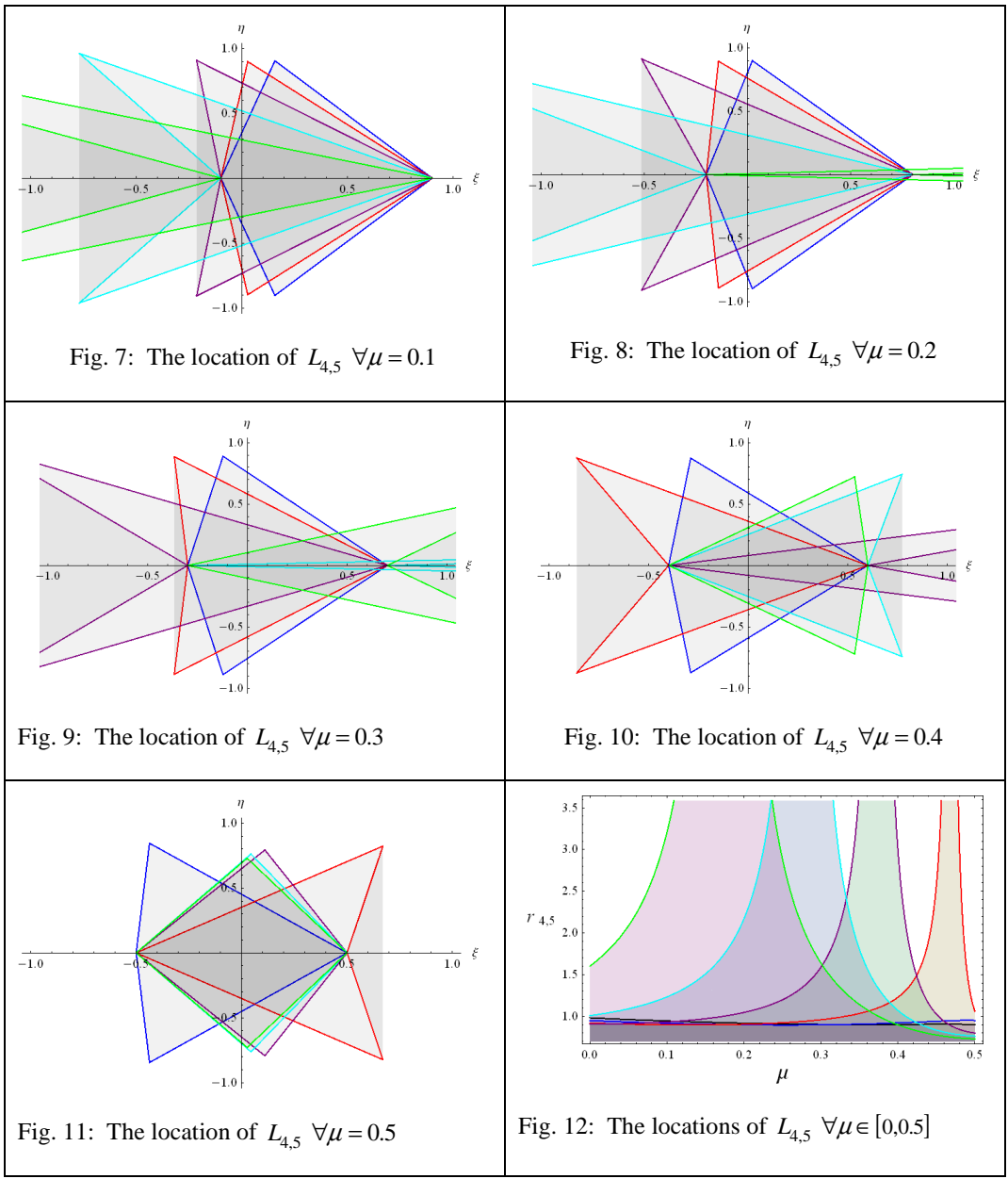

Fig. 7: The location of $L_{4,5}$ $\forall \mu = 0.1$

Fig. 8: The location of $L_{4,5}$ $\forall \mu = 0.2$

Fig. 9: The location of $L_{4,5}$ $\forall \mu = 0.3$

Fig. 10: The location of $L_{4,5}$ $\forall \mu = 0.4$

Fig. 11: The location of $L_{4,5}$ $\forall \mu = 0.5$

Fig. 12: The locations of $L_{4,5}$ $\forall \mu \in [0, 0.5]$

### 3.3. Analysis of different eccentricities on $L_{4,5}$

On Figs. 13-18, we plotted The location of $L_{4,5}$ points in ERTBP, assuming constant nebular and PR drag. we applied our case study for constant radiant primaries. We consider a wide range of eccentricities $e$ for different mass ratios. The curve colors: blue, red. purple, cyan, green, orange, pink, yellow and brown indicate the cases from $e=0.1$ to $e=0.9$ respectively. Fig. 13. shows the eccentricity effect on the triangular points $L_{4,5}$ for systems with $\mu=0.1$. As is clear from the figure, the higher the eccentricity the larger the shift in $L_{4,5}$ location. It seems that the case with $e=0.9, \mu=0.1$ does not appear in the figure, it may not represent a natural system of RTBP. In Figs. 13, 14, 15, 16 some cases exist in which we cannot properly conclude the existence of the triangular equilibrium points, or at least there exists but very far from the barycenter, see the discontinuity of the yellow curve, i.e for $e \geq 0.8, \mu=0.1$ in Fig. 13, orange, pink, yellow and brown, i.e $e \geq 0.6, \mu=0.2$ in Fig. 14, green, orange, pink, yellow and brown curves, i.e. $e \geq 0.5, \mu=0.3$ in Fig. 14, and purple, cyan, green, orange, pink, yellow and brown, i.e. $e \geq 0.3, \mu=0.4$ in Fig. 14. Fig. 17 shows the locations of $L_{4,5}$ for the a systems of equal masses with different given eccentricities. It seems that all cases are represented and all cases have triangular equilibrium points lie to the right of the $\eta$-axis All five cases are represented in the rectangular Fig. 18.

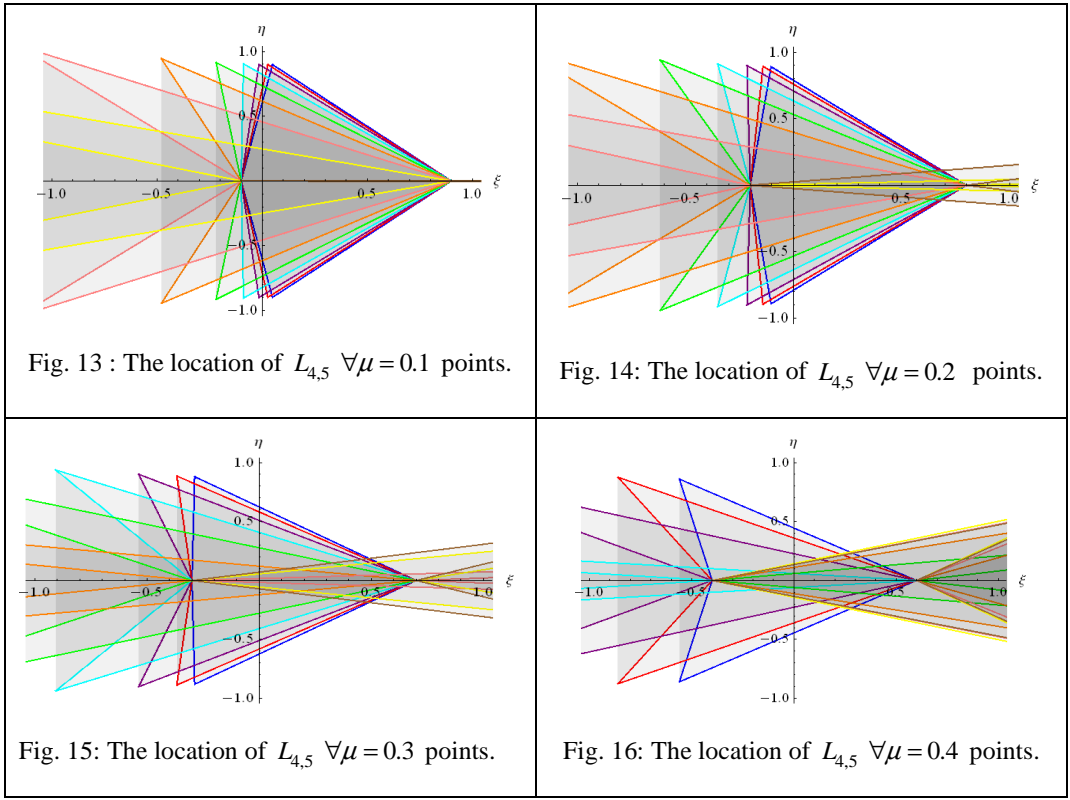

Fig. 13 : The location of $L_{4,5}\ \forall \mu=0.1$ points.

Fig. 14: The location of $L_{4,5}\ \forall \mu=0.2$ points.

Fig. 15: The location of $L_{4,5}\ \forall \mu=0.3$ points.

Fig. 16: The location of $L_{4,5}\ \forall \mu=0.4$ points.

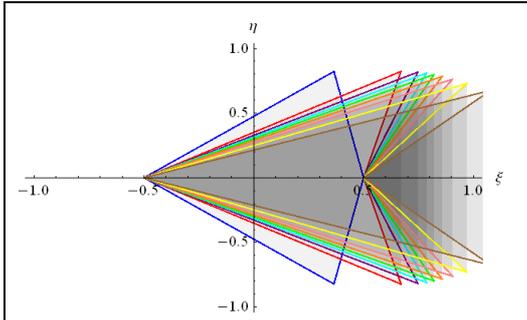
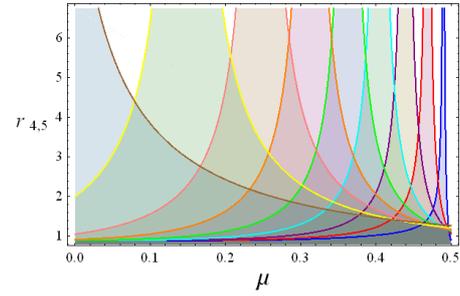

Fig. 17: The location of $L_{4,5}$ $\forall \mu = 0.5$ points.

Fig. 18: The locations of $L_{4,5}$ $\forall \mu \in [0, 0.5]$.

### 3.4. Analysis of drag effect on $L_{4,5}$

On Figs. 19-24, we plotted The location of $L_{4,5}$ points in ERTBP, assuming constant PR drag and constant phtogravitational process. we applied our case study for constant eccentricity of primaries orbit $e = 0.2$. We considered some values of nebular drag for different mass ratios. The curve colors: blue, red. purple, cyan, green, orange, pink, yellow and brown indicate the cases from $\rho = 0.1$ to $\rho = 0.9$ respectively. Fig. 19. shows the effect is so small and all equilibrium points lie to the right of $\eta$-axis while in Fig. 20 the situation is conversed. The remaining of analysis is similar to that proposed in section 6.

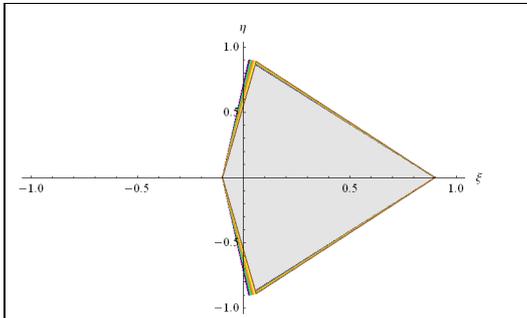
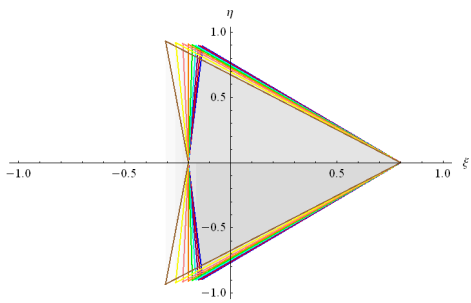

Fig. 19: The location of $L_{4,5}$ $\forall \mu = 0.1$ points.

Fig. 20: The location of $L_{4,5}$ $\forall \mu = 0.2$ points.

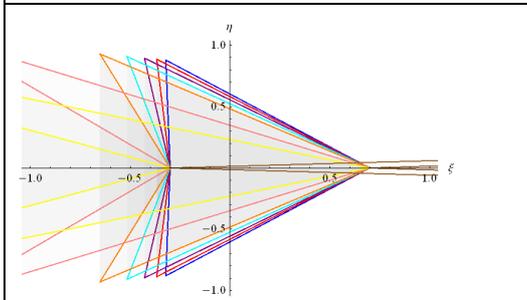
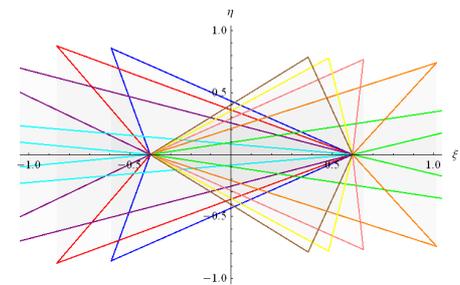

Fig. 21: The location of $L_{4,5}$ $\forall \mu = 0.3$ points.

Fig. 22: The location of $L_{4,5}$ $\forall \mu = 0.4$ points.

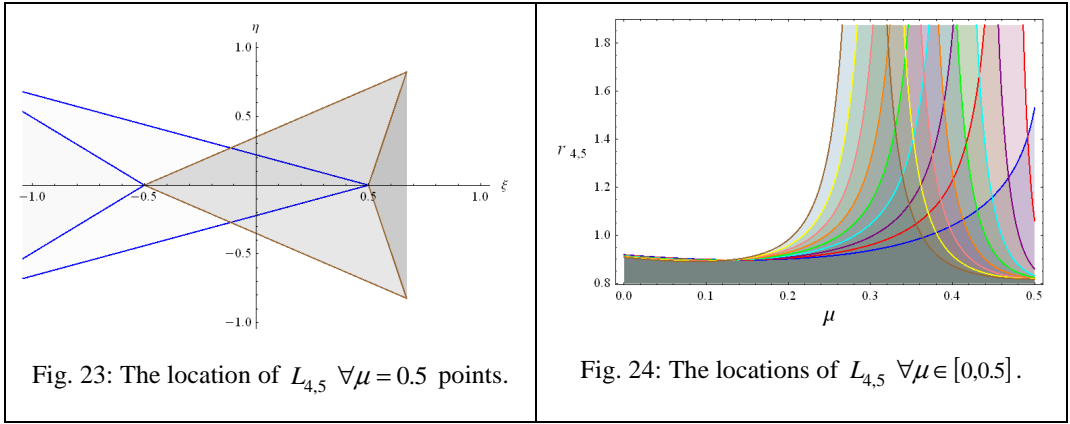

Fig. 23: The location of $L_{4,5}$ $\forall \mu = 0.5$ points.  Fig. 24: The locations of $L_{4,5}$ $\forall \mu \in [0,0.5]$.

## 4. Linear stability of the triangular points

The position of the infinitesimal body is displaced a little from the equilibrium point due to the included perturbations. If the resultant motion of the infinitesimal mass is a rapid departure from the vicinity of the point, we can call such a position of equilibrium point an "unstable one", if however the body merely oscillates about the equilibrium point, it is said to be a "stable position" (in the sense of Lyapunov). In order to analyze the stability, one starts by introducing a displacement $(\delta_\xi, \delta_\eta)$ from the libration points, say $\xi = \xi_{L_{4,5}} + \delta_\xi$, $\eta = \eta_{L_{4,5}} + \delta_\eta$, where $(\xi_{L_{4,5}}, \eta_{L_{4,5}})$ coincides with one of the five stationary solutions. Set $\Phi_\sigma = \mathcal{E} U_\sigma + F_\sigma^{PR} + F_\sigma^{NG}$, $\sigma \equiv \xi, \eta$, then, the linearized equations near the triangular point can be written as

$$\left. \begin{array}{l} \delta_\xi'' - 2\delta_\eta' = \Phi_{\xi\xi}^{(L_{4,5})}\delta_\xi + \Phi_{\xi\eta}^{(L_{4,5})}\delta_\eta, \\ \delta_\eta'' + 2\delta_\xi' = \Phi_{\eta\xi}^{(L_{4,5})}\delta_\xi + \Phi_{\eta\eta}^{(L_{4,5})}\delta_\eta \end{array} \right\} \quad (17)$$

$\Phi_{\xi\xi}^{(L_{4,5})}$ denotes the second derivative of $\Phi$ with respect to $\xi$ computed at the stationary solution $(\xi_{L_{4,5}}, \eta_{L_{4,5}})$ (similarly for the other derivatives).

The characteristic equation corresponding to (16), is

$$\lambda^4 - \left(\Phi_{\xi\xi}^{(L_{4,5})} + \Phi_{\eta\eta}^{(L_{4,5})} - 4\right)\lambda^2 + \Phi_{\xi\xi}^{(L_{4,5})}\Phi_{\eta\eta}^{(L_{4,5})} - \Phi_{\xi\eta}^{(L_{4,5})}\Phi_{\eta\xi}^{(L_{4,5})} = 0 \quad (18)$$

First compute the partial derivatives required for equation (18) as

$$\Phi_{\xi\xi} = \mathcal{E}\left[n^2 - (1-\mu)\left(\frac{1}{r_1^3} + \frac{3}{r_1^5}(\xi+\mu)^2\right) - (1-\mu)A_\sigma\left(\frac{3}{2r_1^5} + \frac{15}{2r_1^7}(\xi+\mu)^2\right)\right.$$

$$-\mu\left(\frac{1}{r_2^3}+\frac{3}{r_2^5}(\xi+\mu-1)^2\right)-\mu A_\gamma\left(\frac{3}{2r_2^5}+\frac{15}{2r_2^7}(\xi+\mu-1)^2\right)\Bigg]$$

$$+k_{PR}\frac{\xi(\xi+\mu)\eta}{r_1^3}+k_{NG}\left(1-\Omega_g\right)^2\rho\frac{\xi(\xi+\mu)\eta}{r}=0 \qquad (19)$$

$$\Phi_{\eta\eta}=\mathcal{E}n^2-\mathcal{E}\Bigg[(1-\mu)\left(\frac{3\eta^2}{r_1^5}+\frac{1}{r_1^3}\right)+3(1-\mu)A_\sigma\left(\frac{5\eta^2}{2r_1^7}+\frac{1}{2r_1^5}\right)$$

$$+\mu\left(\frac{1}{r_2^3}+\frac{\eta^2}{r_2^5}\right)-3\mu A_\gamma\left(\frac{1}{2r_2^5}+\frac{\eta^2}{2r_2^7}\right)\Bigg]$$

$$+k_{PR}\frac{\xi\eta}{r_1^3}+k_{NG}\left(1-\Omega_g\right)^2\rho\frac{\xi\eta}{r}=0 \qquad (20)$$

$$\Phi_{\eta\xi}=\mathcal{E}\eta\left[\frac{3(1-\mu)}{r_1^5}(\xi+\mu)+\frac{3(1-\mu)A_\sigma}{2r_1^7}(\xi+\mu)+\frac{\mu}{r_2^5}(\xi+\mu-1)+\frac{3\mu A_\gamma}{2r_2^7}(\xi+\mu-1)\right]$$

$$-k_{PR}\left[\frac{1}{r_1^2}-\frac{\eta^2}{r_1^4}\right]-k_{NG}\left(1-\Omega_g\right)^2\rho\left[\frac{\eta^2}{r_1}+r\right]=0 \qquad (21)$$

$$\Phi_{\eta\xi}=\mathcal{E}\eta\left[\frac{3(1-\mu)}{r_1^5}(\xi+\mu)+\frac{3(1-\mu)A_\sigma}{2r_1^7}(\xi+\mu)+\frac{\mu}{r_2^5}(\xi+\mu-1)+\frac{3\mu A_\gamma}{2r_2^7}(\xi+\mu-1)\right]$$

$$+k_{PR}\left[\frac{1}{r_1^2}-\frac{\xi(\xi+\mu)}{r_1^4}\right]-k_{NG}\left(1-\Omega_g\right)^2\rho\left[\frac{\xi(\xi+\mu)}{r_1}+r\right]=0 \qquad (22)$$

Let us suppose that $\lambda=i\omega$, so equation (18) can be write in the form

$$\omega^4-\mathcal{U}_2^{L_{4,5}}\omega^2+\mathcal{U}_0^{L_{4,5}}=0 \qquad (23)$$

With

$$\mathcal{U}_2^{(L_{4,5})}=-\left(\Phi_{\xi\xi}^{(L_{4,5})}+\Phi_{\eta\eta}^{(L_{4,5})}-4\right), \qquad \mathcal{U}_0^{L_4}=\Phi_{\xi\xi}^{(L_{4,5})}\Phi_{\eta\eta}^{(L_{4,5})}-\Phi_{\xi\eta}^{(L_{4,5})}\Phi_{\eta\xi}^{(L_{4,5})}$$

The required partial derivatives are obtained using equation (19) - (22) after setting the locations of $L_4$ and $L_5$ as given by $r_1=1+\varepsilon_1$ and $r_2=1+\varepsilon_2$, we have

$$\Phi_{\xi\xi}^{(L_{4,5})}=n^2-(1-\mu)\left((1-3\varepsilon_1)-3(1-5\varepsilon_1)(\xi_{4,5}+\mu)^2\right)$$

$$-(1-\mu)A_\sigma\left(\frac{3}{2}(1-5\varepsilon_1)+\frac{15}{2}(1-7\varepsilon_1)(\xi_{4,5}+\mu)^2\right)$$

$$-\mu\left((1-3\varepsilon_2)-3(1-5\varepsilon_2)(\xi_{4,5}+\mu-1)^2\right)$$

$$-\mu A_\gamma\left(\frac{3}{2}(1-5\varepsilon_2)+\frac{15}{2}(1-7\varepsilon_2)(\xi_{4,5}+\mu-1)^2\right)$$

$$+k_{PR}\xi_{4,5}\left(\xi_{4,5}+\mu\right)\eta_{4,5}(1-3\varepsilon_1)+k_{NG}\left(1-\Omega_g\right)^2\rho\frac{\xi_{4,5}\left(\xi_{4,5}+\mu\right)\eta_{4,5}}{(\xi_{4,5}^2+\eta_{4,5}^2)^{1/2}}=0 \qquad (24)$$

$$\Phi_{\eta\eta}^{(L_{4,5})}=n^2-\left[(1-\mu)\left((1-3\varepsilon_1)+3\eta_{4,5}^2(1-5\varepsilon_1)\right)+3(1-\mu)A_\sigma\left(\frac{5\eta_{4,5}^2}{2}(1-7\varepsilon_1)+\frac{1}{2}(1-5\varepsilon_1)\right)\right.$$

$$\left.+\mu\left((1-3\varepsilon_2)+3\eta_{4,5}^2(1-5\varepsilon_2)\right)-3\mu A_\gamma\left(\frac{1}{2}(1-5\varepsilon_2)+\frac{\eta_{4,5}^2}{2}(1-7\varepsilon_2)\right)\right]$$

$$+k_{PR}\xi_{4,5}\eta_{4,5}(1-3\varepsilon_1)+k_{NG}\left(1-\Omega_g\right)^2\rho\frac{\xi_{4,5}\eta_{4,5}}{(\xi_{4,5}^2+\eta_{4,5}^2)^{1/2}}=0 \qquad (25)$$

$$\Phi_{\xi\eta}^{(L_{4,5})}=3\eta_{4,5}\left[(1-\mu)(1-5\varepsilon_1)(\xi_{4,5}+\mu)+\frac{(1-\mu)A_\sigma}{2}(1-7\varepsilon_1)(\xi_{4,5}+\mu)\right.$$

$$\left.+\mu(1-5\varepsilon_2)(\xi_{4,5}+\mu-1)+\frac{\mu A_\gamma}{2}(1-5\varepsilon_2)(\xi+\mu-1)\right]$$

$$-k_{PR}\left[(1-2\varepsilon_1)-\eta_{4,5}^2(1-4\varepsilon_1)\right]$$

$$-k_{NG}\left(1-\Omega_g\right)^2\rho\left[\eta_{4,5}^2(1-\varepsilon_1)+(\xi_{4,5}^2+\eta_{4,5}^2)^{1/2}\right]=0 \qquad (26)$$

$$\Phi_{\eta\xi}^{(L_{4,5})}=3\eta_{4,5}\left[(1-\mu)(1-5\varepsilon_1)(\xi_{4,5}+\mu)+\frac{(1-\mu)A_\sigma}{2}(1-7\varepsilon_1)(\xi_{4,5}+\mu)\right.$$

$$\left.+\mu(1-5\varepsilon_2)(\xi_{4,5}+\mu-1)+\frac{\mu A_\gamma}{2}(1-5\varepsilon_2)(\xi+\mu-1)\right]$$

$$+k_{PR}\left[(1-2\varepsilon_1)-\xi_{4,5}\left(\xi_{4,5}+\mu\right)(1-4\varepsilon_1)\right]$$

$$-k_{NG}\left(1-\Omega_g\right)^2\rho\left[\xi_{4,5}\left(\xi_{4,5}+\mu\right)(1-\varepsilon_1)+(\xi_{4,5}^2+\eta_{4,5}^2)^{1/2}\right]=0 \qquad (27)$$

As a fast check, removing all perturbations one gets;

$$\Phi_{\xi\xi}^{(L_{4,5})}=\frac{3}{4}, \qquad \Phi_{\eta\eta}^{(L_{4,5})}=\frac{9}{4}, \qquad \Phi_{\xi\eta}^{(L_{4,5})}=\frac{3\sqrt{3}}{4}(1-2\mu)$$

$$\mathcal{U}_2^{(L_{4,5})}=1, \qquad\qquad\qquad \mathcal{U}_0^{(L_{4,5})}=\frac{27}{4}\mu-\frac{27}{4}\mu^2 \qquad (28)$$

### 4.1. Graphical representations and analysis

In the following figures from Fig. 25 to Fig. 32, we will plotted the stability/instability regions. The roots $\omega_1, \omega_2, \omega_3$ and $\omega_4$ are plotted against different values of mass ratios for different types of perturbations. In Fig. 25 we plotted six colored curves namely the blue, red, purple, cyan, green and orange corresponds to six different values of eccentricities, from $e = 0.1$, to $e = 0.6$ respectively. The Fig. 25 revealed that we have stability regions for $\mu = [0, 0.1]$. The change in stability with eccentricity in this mentioned domain seems to be nonlinear. In Fig. 26 we plotted five colored curves namely the blue, red, purple, cyan and green corresponds to five different values of eccentricities, from $e = 0.705$, to $e = 0.66$ respectively. The Fig. 26 revealed that we have two stability regions for $\mu = [0, 0.185] \cup [0.262, 0.5]$. We have instability region $\mu = (0.185, 0.262)$. It seems that increasing the eccentricity the enlarge the stability regions and vice versa. While in Fig. 27 we have plotted the stability/instability regions for a new set of eccentricities $e \in [0.71, 0.79]$. The stability regions are merged to one region corresponds to $\mu = [0, 0.36]$. Again when we consider a very high eccentricity $e \in [0.9, 0.908]$, see Fig. 28, we have stability region for the whole domain of the mass ratio except in the neighborhood of $\mu = 0.3$. In Fig. 29, and Fig. 30, we have plotted the photogravitational effects correspond to $q \in [0.9, 0.6]$ and $q \in [0.55, 0.51]$ respectively on the stability regions of the multi-dissipative photogravitational ERTBP. We have stability/asymptotic/instability regions, see Fig. 30. In Fig. 31, we have plotted ten curves to visualize the effect of the nebular gas (Stokes) drag with density $\rho \in [0.1, 0.9]$ on the stability regions multi-dissipative ERTBP. While Fig. 32: The triaxial effect with coefficients $A_\sigma, A_\gamma = 10^{-3}$ to $A_\sigma, A_\gamma = 10^{-2}$ on the stability regions multi-dissipative ERTBP. In all stability figures we have only two roots that cause the stability, except for the very high eccentricity we have the four roots that cause the stability.

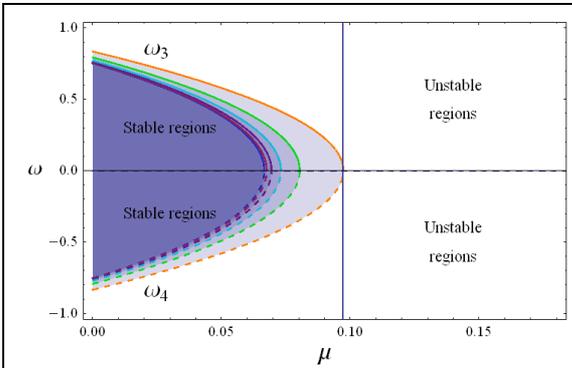

Fig. 25: The effect of the eccentricity $e = 0.1$ to

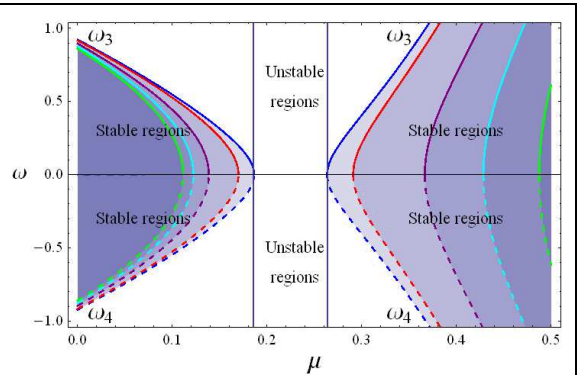

Fig. 26: The effect of the eccentricity $e = 0.64$ to

| $e = 0.6$ on the stability regions multi-dissipative photogravitational ERTBP. | $e = 0.705$ on the stability regions multi-dissipative photogravitational ERTBP. |
|---|---|
| 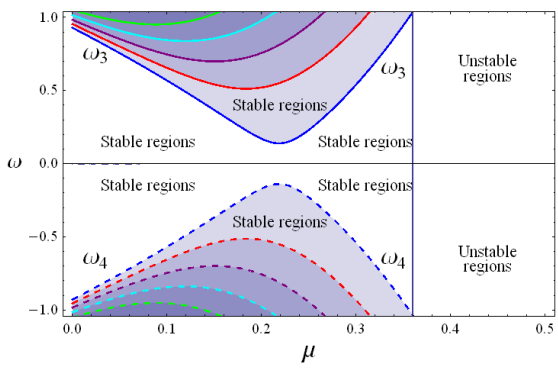<br>Fig. 27: The effect of the eccentricity $e = 0.71$ to $e = 0.79$ on the stability regions multi-dissipative photogravitational ERTBP. | 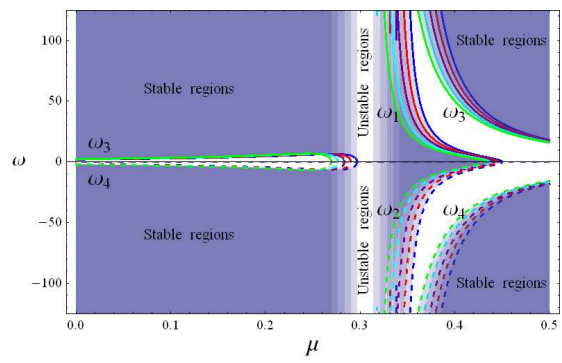<br>Fig. 28: The effect of the eccentricity $e = 0.9$ to $e = 0.908$ on the stability regions multi-dissipative photogravitational ERTBP. |
| 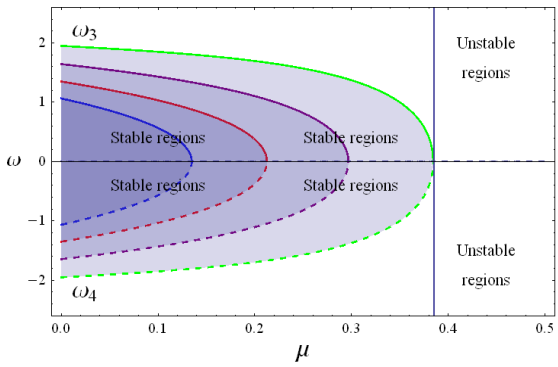<br>Fig. 29: The photogravitational effect $q = 0.9$ to $e = 0.6$ on the stability regions multi-dissipative photogravitational ERTBP. | 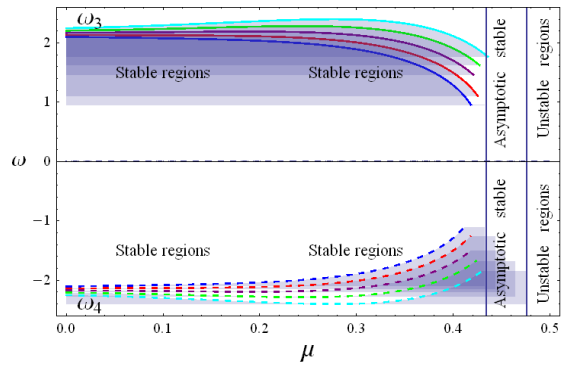<br>Fig. 30: The effect of the photogravitational $q = 0.55$ to $q = 0.51$ on the stability regions multi-dissipative photogravitational ERTBP. |
| 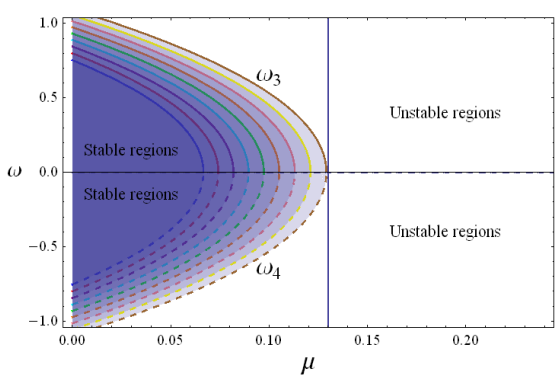<br>Fig. 31: The effect of the nebular gas (Stokes) drag with density $\rho = 0.1$ to $\rho = 0.9$ on the stability regions multi-dissipative ERTBP. | 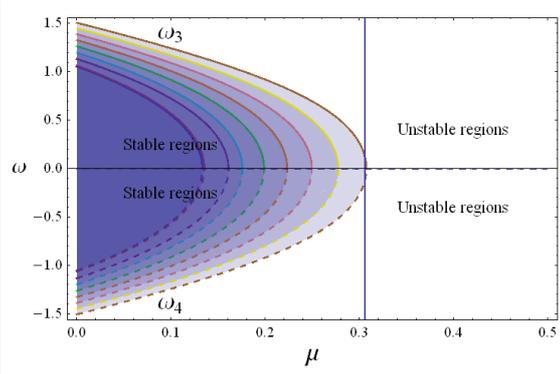<br>Fig. 32: The triaxial effect with coefficients $A_\sigma, A_\gamma = 10^{-3}$ to $A_\sigma, A_\gamma = 10^{-2}$ on the stability regions multi-dissipative ERTBP. |

## 5. Conclusion

We treated the multi-dissipative elliptic restricted three body problem. The primaries are assumed oblate and triaxial as well as radiant sources. We have computed the perturbed locations of triangular points. We investigated the stability of the triangular points under changing one or more perturbing parameter, namly; the eccentricity of the primaries' orbits, oblateness coefficients and the triaxial parameters of the primaries, the photogravitational effects and the drag perturbations. We can conclude our remarkable results as follows:-

At certain values of the considered perturbing parameters, we haven't triangular equilibrium points, as some curves are not continuous. The bigger the mass loss the larger the shift in $L_{4,5}$ location, it seems that the case with $q=0.5$, $\mu=0.1$ has no triangular equilibrium point, or at least there exists but very far from the origin. The higher the eccentricity the larger the shift in $L_{4,5}$ location. The change in the stability/instability regions with eccentricity seems to be nonlinear. It seems that increasing the eccentricity the enlarge the stability regions and vice versa. It is revealed that we have two disjoint stability regions. The stability/instability regions for a new set of eccentricities $e \in [0.71, 0.79]$ are merged to one region corresponds to $\mu = [0, 0.36]$. Again when we consider a very high eccentricity $e \in [0.9, 0.908]$, we have stability region for the whole domain of the mass ratio except in the neighborhood of $\mu = 0.3$. We have obtained stability/asymptotic/instability regions due to photogravitational effects correspond to $q \in [0.55, 0.51]$.

## 6. Appendix A

The non-vanishing coefficients $\mathcal{D}_{0,k}$, $\mathcal{D}_{1,k}$ and $\mathcal{D}_{2,k}$ are given by

$$\mathcal{D}_{0,0} = \frac{2}{\sqrt{3}}n^4 - \frac{5}{\sqrt{3}}n^2 q - 2\sqrt{3}q^2 - \frac{7\sqrt{3}}{2}n^2 q A_\sigma - 8\sqrt{3}q^2 A_\sigma - \frac{15\sqrt{3}}{2}q^2 A_\sigma^2 - 2n^2 \rho k_{NG}$$

$$-\frac{3}{2}q\rho k_{NG} - \frac{3\sqrt{3}}{2}q\rho k_{NG} - n^2 k_{PR} + \frac{\sqrt{3}}{3}n^2 k_{PR} - \frac{1}{2}q k_{PR} - \frac{11\sqrt{3}}{6}q k_{PR} - \frac{9}{4}q A_\sigma k_{PR}$$

$$-\frac{17\sqrt{3}}{4}q A_\sigma k_{PR} + 4n^2 \rho k_{NG}\Omega_g + 3q\rho k_{NG}\Omega_g + 3\sqrt{3}q\rho k_{NG}\Omega_g - 3\mu\rho k_{NG}\Omega_g$$

$$+ 3\sqrt{3}\mu\rho k_{NG}\Omega_g - 6n^2 \mu\rho k_{NG}\Omega_g + \frac{15}{2}q\rho A_\sigma k_{NG}\Omega_g + \frac{15\sqrt{3}}{2}q\rho A_\sigma k_{NG}\Omega_g$$

$$- 2n^2 \rho k_{NG}\Omega_g^2 - \frac{3}{2}q\rho k_{NG}\Omega_g^2 - \frac{3\sqrt{3}}{2}q\rho k_{NG}\Omega_g^2 - \frac{15}{4}q\rho A_\sigma k_{NG}\Omega_g^2$$

$$- \frac{15\sqrt{3}}{4}q\rho A_\sigma k_{NG}\Omega_g^2 - \frac{15}{4}q\rho A_\sigma k_{NG} - \frac{15\sqrt{3}}{4}q\rho A_\sigma k_{NG}$$

$$\mathcal{D}_{0,1} = \frac{3\sqrt{3}}{2}q + \frac{5}{\sqrt{3}}n^2 q + 4\sqrt{3}q^2 + \frac{7\sqrt{3}}{2}n^2 A_\gamma + \frac{7\sqrt{3}}{2}qA_\gamma + \frac{19\sqrt{3}}{4}qA_\sigma$$

$$+ \frac{7\sqrt{3}}{2}n^2 qA_\sigma + 16\sqrt{3}q^2 A_\sigma + \frac{15\sqrt{3}}{4}qA_\gamma A_\sigma + 15\sqrt{3}q^2 A_\sigma^2 + \frac{3}{2}\rho k_{NG}$$

$$- \frac{3\sqrt{3}}{2}\rho k_{NG} + 3n^2 \rho k_{NG} + \frac{15}{4}q\rho k_{NG} + \frac{15\sqrt{3}}{4}q\rho k_{NG} + \frac{15}{4}\rho A_\gamma k_{NG}$$

$$+ \frac{15\sqrt{3}}{4}\rho A_\gamma k_{NG} + \frac{75}{8}q\rho A_\sigma k_{NG} + \frac{75\sqrt{3}}{8}q\rho A_\sigma k_{NG} + k_{PR} - \frac{1}{\sqrt{3}}k_{PR}$$

$$- 2n^2 k_{PR} - \frac{2}{\sqrt{3}}n^2 k_{PR} - \frac{3}{2}qk_{PR} + \frac{5\sqrt{3}}{2}qk_{PR} + \frac{21}{4}A_\gamma k_{PR} + \frac{17\sqrt{3}}{4}A_\gamma k_{PR}$$

$$+ \frac{1}{\sqrt{3}}n^2 - \frac{3}{4}qA_\sigma k_{PR} + \frac{21\sqrt{3}}{4}qA_\sigma k_{PR} - \frac{15}{2}q\rho k_{NG}\Omega_g - \frac{15\sqrt{3}}{2}q\rho k_{NG}\Omega_g$$

$$- \frac{15}{2}\rho A_\gamma k_{NG}\Omega_g - \frac{15\sqrt{3}}{2}\rho A_\gamma k_{NG}\Omega_g - \frac{75}{4}q\rho A_\sigma k_{NG}\Omega_g - \frac{75\sqrt{3}}{4}q\rho A_\sigma k_{NG}\Omega_g$$

$$+ \frac{3}{2}\rho k_{NG}\Omega_g^2 - \frac{3\sqrt{3}}{2}\rho k_{NG}\Omega_g^2 + 3n^2 \rho k_{NG}\Omega_g^2 + \frac{15}{4}q\rho k_{NG}\Omega_g^2 + \frac{15\sqrt{3}}{4}q\rho k_{NG}\Omega_g^2$$

$$+ \frac{15}{4}\rho A_\gamma k_{NG}\Omega_g^2 + \frac{15\sqrt{3}}{4}\rho A_\gamma k_{NG}\Omega_g^2 + \frac{75}{8}q\rho A_\sigma k_{NG}\Omega_g^2 + \frac{75\sqrt{3}}{8}q\rho A_\sigma k_{NG}\Omega_g^2$$

$$\mathcal{D}_{0,2} = -\sqrt{3} - \frac{3\sqrt{3}}{2}q - 2\sqrt{3}q^2 - \frac{37\sqrt{3}}{4}A_\gamma - \frac{7\sqrt{3}}{2}qA_\gamma + \frac{15\sqrt{3}}{4}A_\gamma^2$$

$$- \frac{19\sqrt{3}}{4}qA_\sigma - 8\sqrt{3}q^2 A_\sigma - \frac{15\sqrt{3}}{4}qA_\gamma A_\sigma - \frac{15\sqrt{3}}{2}q^2 A_\sigma^2 - \frac{9}{4}\rho k_{NG}$$

$$+ \frac{9\sqrt{3}}{4}\rho k_{NG} - 4n^2 \rho k_{NG} - \frac{21}{4}q\rho k_{NG} - \frac{21\sqrt{3}}{4}q\rho k_{NG} - \frac{45}{8}\rho A_\gamma k_{NG}$$

$$- \frac{45\sqrt{3}}{8}\rho A_\gamma k_{NG} - \frac{105}{8}q\rho A_\sigma k_{NG} - \frac{105\sqrt{3}}{8}q\rho A_\sigma k_{NG} + k_{PR} - \frac{7}{\sqrt{3}}\mu^2 k_{PR}$$

$$+ 2qk_{PR} - \frac{2}{\sqrt{3}}qk_{PR} - 3A_\gamma k_{PR} - \sqrt{3}A_\gamma k_{PR} + 3qA_\sigma k_{PR} - \sqrt{3}qA_\sigma k_{PR}$$

$$+ \frac{9}{2}\rho k_{NG}\Omega_g - \frac{9\sqrt{3}}{2}\rho k_{NG}\Omega_g + 8n^2 \rho k_{NG}\Omega_g + \frac{21}{2}q\rho k_{NG}\Omega_g$$

$$+ \frac{21\sqrt{3}}{2}q\rho k_{NG}\Omega_g + \frac{45}{4}\rho A_\gamma k_{NG}\Omega_g + \frac{45\sqrt{3}}{4}\rho A_\gamma k_{NG}\Omega_g$$

$$+ \frac{105}{4}q\rho A_\sigma k_{NG}\Omega_g + \frac{105\sqrt{3}}{4}q\rho A_\sigma k_{NG}\Omega_g - \frac{9}{4}\rho k_{NG}\Omega_g^2$$

$$+ \frac{9\sqrt{3}}{4}\rho k_{NG}\Omega_g^2 - 4n^2 \rho k_{NG}\Omega_g^2 - \frac{21}{4}q\rho k_{NG}\Omega_g^2 - \frac{21\sqrt{3}}{4}q\rho k_{NG}\Omega_g^2$$

$$-\frac{45}{8}\rho A_\gamma k_{NG}\Omega_g^2 - \frac{45\sqrt{3}}{8}\rho A_\gamma k_{NG}\Omega_g^2 - \frac{105}{8}q\rho A_\sigma k_{NG}\Omega_g^2 - \frac{105\sqrt{3}}{8}q\rho A_\sigma k_{NG}\Omega_g^2$$

$$\mathcal{D}_{0,3} = 3\rho k_{NG} - 3\sqrt{3}\rho k_{NG} + 3q\rho k_{NG} + 3\sqrt{3}q\rho k_{NG} + \frac{15}{2}\rho A_\gamma k_{NG} + \frac{15}{2}\sqrt{3}\rho A_\gamma k_{NG}$$
$$+ \frac{15}{2}q\rho A_\sigma k_{NG} + \frac{15}{2}\sqrt{3}q\rho A_\sigma k_{NG} - 6\rho k_{NG}\Omega_g + 6\sqrt{3}\rho k_{NG}\Omega_g - 6q\rho k_{NG}\Omega_g$$
$$- 6\sqrt{3}q\rho k_{NG}\Omega_g - 15\rho A_\gamma k_{NG}\Omega_g - 15\sqrt{3}\rho A_\gamma k_{NG}\Omega_g - 15q\rho A_\sigma k_{NG}\Omega_g$$
$$- 15\sqrt{3}q\rho A_\sigma k_{NG}\Omega_g + 3\rho k_{NG}\Omega_g^2 - 3\sqrt{3}\rho k_{NG}\Omega_g^2 + 3q\rho k_{NG}\Omega_g^2 + 3\sqrt{3}q\rho k_{NG}\Omega_g^2$$
$$+ \frac{15}{2}\rho A_\gamma k_{NG}\Omega_g^2 + \frac{15\sqrt{3}}{2}\rho A_\gamma k_{NG}\Omega_g^2 + \frac{15}{2}q\rho A_\sigma k_{NG}\Omega_g^2 + \frac{15\sqrt{3}}{2}q\rho A_\sigma k_{NG}\Omega_g^2$$

$$\mathcal{D}_{1,0} = \frac{n^4}{2\sqrt{3}} - \frac{\sqrt{3}n^4}{2} - \frac{n^2 q}{\sqrt{3}} + \frac{q^2}{2\sqrt{3}} + \frac{\sqrt{3}q^2}{2} - \frac{1}{2}\sqrt{3}n^2 q A_\sigma + 2\sqrt{3}q^2 A_\sigma + \frac{3}{2}\sqrt{3}q^2 A_\sigma^2$$
$$- n^2\rho k_{NG} - \frac{n^2\rho k_{NG}}{2\sqrt{3}} - \frac{1}{2}\sqrt{3}n^2\rho k_{NG} + \frac{q\rho k_{NG}}{2\sqrt{3}} + \frac{1}{2}\sqrt{3}q\rho k_{NG} + \sqrt{3}q\rho A_\sigma k_{NG} - n^2 k_{PR}$$
$$- \frac{n^2 k_{PR}}{2\sqrt{3}} - \frac{1}{2}\sqrt{3}n^2 k_{PR} + \frac{q k_{PR}}{2\sqrt{3}} + \frac{1}{2}\sqrt{3}q k_{PR} + \sqrt{3}q A_\sigma k_{PR} + 2n^2\rho k_{NG}\Omega_g + \frac{n^2\rho k_{NG}\Omega_g}{\sqrt{3}}$$
$$+ \sqrt{3}n^2\rho k_{NG}\Omega_g - \frac{q\rho k_{NG}\Omega_g}{\sqrt{3}} - \sqrt{3}q\rho k_{NG}\Omega_g - 2\sqrt{3}q\rho A_\sigma k_{NG}\Omega_g - n^2\rho k_{NG}\Omega_g^2$$
$$- \frac{n^2\rho k_{NG}\Omega_g^2}{2\sqrt{3}} - \frac{1}{2}\sqrt{3}n^2\rho k_{NG}\Omega_g^2 + \frac{q\rho k_{NG}\Omega_g^2}{2\sqrt{3}} + \frac{1}{2}\sqrt{3}q\rho k_{NG}\Omega_g^2 + \sqrt{3}q\rho A_\sigma k_{NG}\Omega_g^2$$

$$\mathcal{D}_{1,1} = \frac{3\sqrt{3}}{2}n^2 - \frac{1}{\sqrt{3}}n^4 - \frac{\sqrt{3}}{2}q + \frac{2}{\sqrt{3}}n^2 q - \frac{1}{\sqrt{3}}q^2 - \sqrt{3}q^2 - \sqrt{3}n^2 A_\gamma - \frac{\sqrt{3}}{2}q A_\gamma$$
$$- \frac{3\sqrt{3}}{4}q A_\sigma + \sqrt{3}n^2 q A_\sigma - 4\sqrt{3}q^2 A_\sigma - \frac{3\sqrt{3}}{4}q A_\gamma A_\sigma - 3\sqrt{3}q^2 A_\sigma^2 - \frac{1}{4}\rho k_{NG}$$
$$- \frac{7\sqrt{3}}{12}\rho k_{NG} + \frac{\sqrt{3}}{2}\rho k_{NG} + \frac{13}{4}n^2\rho k_{NG} + \frac{5\sqrt{3}}{4}n^2\rho k_{NG} + \frac{3}{4}q\rho k_{NG} - \frac{1}{2\sqrt{3}}q\rho k_{NG}$$
$$- \frac{7\sqrt{3}}{4}q\rho k_{NG} - \frac{3}{2}\rho A_\gamma k_{NG} - \sqrt{3}\rho A_\gamma k_{NG} + \frac{9}{8}q\rho A_\sigma k_{NG} - \frac{23\sqrt{3}}{8}q\rho A_\sigma k_{NG}$$
$$- \sqrt{3}q k_{PR} - \frac{\sqrt{3}}{6}q\rho k_{NG}\Omega_g^2 - \frac{1}{4}k_{PR} - \frac{7\sqrt{3}}{12}k_{PR} + \frac{\sqrt{3}}{2}k_{PR} + 2n^2 k_{PR} + q k_{PR}$$
$$+ \frac{1}{\sqrt{3}}n^2 k_{PR} - \frac{3}{2}A_\gamma k_{PR} - \sqrt{3}A_\gamma k_{PR} + \frac{3}{2}q A_\sigma k_{PR} - \frac{3\sqrt{3}}{2}q A_\sigma k_{PR} + \frac{1}{2}\rho k_{NG}\Omega_g$$
$$+ \frac{7\sqrt{3}}{6}\rho k_{NG}\Omega_g - \sqrt{3}\rho k_{NG}\Omega_g - \frac{13}{2}n^2\rho k_{NG}\Omega_g - \frac{5\sqrt{3}}{2}n^2\rho k_{NG}\Omega_g - \frac{3}{2}q\rho k_{NG}\Omega_g$$

$$+ \frac{\sqrt{3}}{3} q\rho k_{NG}\Omega_g + \frac{7\sqrt{3}}{2} q\rho k_{NG}\Omega_g + 3\rho A_\gamma k_{NG}\Omega_g + 2\sqrt{3}\rho A_\gamma k_{NG}\Omega_g - \frac{1}{4}\rho k_{NG}\Omega_g^2$$

$$- \frac{7\sqrt{3}}{12}\rho k_{NG}\Omega_g^2 + \frac{\sqrt{3}}{2}\rho k_{NG}\Omega_g^2 - \frac{9}{4}q\rho A_\sigma k_{NG}\Omega_g + \frac{23\sqrt{3}}{4}q\rho A_\sigma k_{NG}\Omega_g$$

$$+ \frac{13}{4}n^2\rho k_{NG}\Omega_g^2 + \frac{5\sqrt{3}}{4}n^2\rho k_{NG}\Omega_g^2 + \frac{3}{4}q\rho k_{NG}\Omega_g^2 - \frac{7\sqrt{3}}{4}q\rho k_{NG}\Omega_g^2$$

$$- \frac{3}{2}\rho A_\gamma k_{NG}\Omega_g^2 + \sqrt{3}\rho A_\gamma k_{NG}\Omega_g^2 + \frac{9}{8}q\rho A_\sigma k_{NG}\Omega_g^2 - \frac{23\sqrt{3}}{8}q\rho A_\sigma k_{NG}\Omega_g^2$$

$$\mathcal{D}_{1,2} = \frac{7}{4\sqrt{3}} - \frac{\sqrt{3}}{4} - \frac{7n^2}{2\sqrt{3}} + \frac{1}{2}\sqrt{3}q - \frac{n^2 q}{\sqrt{3}} + \frac{q^2}{2\sqrt{3}} - \frac{15}{8}q\rho A_\sigma k_{NG}$$

$$+ \frac{1}{2}\sqrt{3}q^2 + \frac{9}{4}\sqrt{3}A_\gamma - \frac{1}{2}\sqrt{3}n^2 A_\gamma + \frac{1}{2}\sqrt{3}qA_\gamma - \frac{3}{4}\sqrt{3}A_\gamma^2$$

$$+ \frac{3}{4}\sqrt{3}qA_\sigma - \frac{1}{2}\sqrt{3}n^2 qA_\sigma + 2\sqrt{3}q^2 A_\sigma + \frac{3}{4}\sqrt{3}qA_\gamma A_\sigma + \frac{3}{2}\sqrt{3}q^2 A_\sigma^2$$

$$+ \sqrt{3}\rho k_{NG} - 4n^2\rho k_{NG} - \frac{3}{2}\sqrt{3}n^2\rho k_{NG} - \frac{5}{4}q\rho k_{NG} + \frac{11}{4}\sqrt{3}q\rho k_{NG}$$

$$+ \frac{27}{8}\rho A_\gamma k_{NG} + \frac{15}{8}\sqrt{3}\rho A_\gamma k_{NG} + \frac{33}{8}\sqrt{3}q\rho A_\sigma k_{NG} - \frac{k_{PR}}{2} + \frac{7k_{PR}}{2\sqrt{3}}$$

$$- qk_{PR} + \frac{qk_{PR}}{\sqrt{3}} + \frac{3}{2}A_\gamma k_{PR} + \frac{1}{2}\sqrt{3}A_\gamma k_{PR} - \frac{3}{2}qA_\sigma k_{PR} + \frac{1}{2}\sqrt{3}qA_\sigma k_{PR}$$

$$- 2\sqrt{3}\rho k_{NG}\Omega_g + 8n^2\rho k_{NG}\Omega_g + 3\sqrt{3}n^2\rho k_{NG}\Omega_g + \frac{5}{2}q\rho k_{NG}\Omega_g$$

$$- \frac{27}{4}\rho A_\gamma k_{NG}\Omega_g - \frac{15}{4}\sqrt{3}\rho A_\gamma k_{NG}\Omega_g + \frac{15}{4}q\rho A_\sigma k_{NG}\Omega_g$$

$$- \frac{33}{4}\sqrt{3}q\rho A_\sigma k_{NG}\Omega_g + \sqrt{3}\rho k_{NG}\Omega_g^2 - 4n^2\rho k_{NG}\Omega_g^2 - \frac{3}{2}\sqrt{3}n^2\rho k_{NG}\Omega_g^2$$

$$- \frac{5}{4}q\rho k_{NG}\Omega_g^2 + \frac{11}{4}\sqrt{3}q\rho k_{NG}\Omega_g^2 + \frac{27}{8}\rho A_\gamma k_{NG}\Omega_g^2 + \frac{15}{8}\sqrt{3}\rho A_\gamma k_{NG}\Omega_g^2$$

$$- \frac{15}{8}q\rho A_\sigma k_{NG}\Omega_g^2 - \frac{33}{8}\sqrt{3}q\rho A_\sigma k_{NG}\Omega_g^2$$

$$\mathcal{D}_{1,3} = -\frac{1}{4}\rho k_{NG} - \frac{3}{4}\sqrt{3}\rho k_{NG} + 3n^2\rho k_{NG} + \frac{n^2\rho k_{NG}}{\sqrt{3}} + \frac{3}{2}q\rho k_{NG}$$

$$- \frac{q\rho k_{NG}}{\sqrt{3}} - \frac{3}{2}\sqrt{3}q\rho k_{NG} - \frac{9}{4}\sqrt{3}\rho A_\gamma k_{NG} + \frac{9}{4}q\rho A_\sigma k_{NG}$$

$$- \frac{11}{4}\sqrt{3}q\rho A_\sigma k_{NG} - \frac{15}{4}\rho A_\gamma k_{NG} + \frac{1}{2}\rho k_{NG}\Omega_g + \frac{3}{2}\sqrt{3}\rho k_{NG}\Omega_g$$

$$- 6n^2\rho k_{NG}\Omega_g - \frac{2n^2\rho k_{NG}\Omega_g}{\sqrt{3}} - 3q\rho k_{NG}\Omega_g + \frac{2q\rho k_{NG}\Omega_g}{\sqrt{3}}$$

$$+3\sqrt{3}q\rho k_{NG}\Omega_g + \frac{15}{2}\rho A_\gamma k_{NG}\Omega_g + \frac{9}{2}\sqrt{3}\rho A_\gamma k_{NG}\Omega_g - \frac{9}{2}q\rho A_\sigma k_{NG}\Omega_g$$

$$+\frac{11}{2}\sqrt{3}q\rho A_\sigma k_{NG}\Omega_g - \frac{1}{4}\rho k_{NG}\Omega_g^2 - \frac{3}{4}\sqrt{3}\rho k_{NG}\Omega_g^2 + 3n^2\rho k_{NG}\Omega_g^2$$

$$+\frac{n^2 \rho k_{NG}\Omega_g^2}{\sqrt{3}} + \frac{3}{2}q\rho k_{NG}\Omega_g^2 - \frac{q\rho k_{NG}\Omega_g^2}{\sqrt{3}} - \frac{3}{2}\sqrt{3}q\rho k_{NG}\Omega_g^2$$

$$-\frac{15}{4}\rho A_\gamma k_{NG}\Omega_g^2 - \frac{9}{4}\sqrt{3}\rho A_\gamma k_{NG}\Omega_g^2 + \frac{9}{4}q\rho A_\sigma k_{NG}\Omega_g^2 - \frac{11}{4}\sqrt{3}q\rho A_\sigma k_{NG}\Omega_g^2$$

$$\mathcal{D}_{1,4} = -\frac{1}{2}\rho k_{NG} + \frac{7\rho k_{NG}}{2\sqrt{3}} - q\rho k_{NG} + \frac{q\rho k_{NG}}{\sqrt{3}} + \frac{3}{2}\rho A_\gamma k_{NG}$$

$$+\frac{1}{2}\sqrt{3}\rho A_\gamma k_{NG} - \frac{3}{2}q\rho A_\sigma k_{NG} + \frac{1}{2}\sqrt{3}q\rho A_\sigma k_{NG} + \rho k_{NG}\Omega_g$$

$$-\frac{7\rho k_{NG}\Omega_g}{\sqrt{3}} + 2q\rho k_{NG}\Omega_g - \frac{2q\rho k_{NG}\Omega_g}{\sqrt{3}} - 3\rho A_\gamma k_{NG}\Omega_g$$

$$-\sqrt{3}\rho A_\gamma k_{NG}\Omega_g + 3q\rho A_\sigma k_{NG}\Omega_g - \sqrt{3}q\rho A_\sigma k_{NG}\Omega_g - \frac{1}{2}\rho k_{NG}\Omega_g^2$$

$$+\frac{7\rho k_{NG}\Omega_g^2}{2\sqrt{3}} - q\rho k_{NG}\Omega_g^2 + \frac{q\rho k_{NG}\Omega_g^2}{\sqrt{3}} + \frac{3}{2}\rho A_\gamma k_{NG}\Omega_g^2 + \frac{1}{2}\sqrt{3}\rho A_\gamma k_{NG}\Omega_g^2$$

$$-\frac{3}{2}q\rho A_\sigma k_{NG}\Omega_g^2 + \frac{1}{2}\sqrt{3}q\rho A_\sigma k_{NG}\Omega_g^2$$

$$\mathcal{D}_{2,0} = -\frac{n^4}{2\sqrt{3}} - \frac{\sqrt{3}n^4}{2} + \frac{13n^2 q}{4\sqrt{3}} + \frac{1}{4}\sqrt{3}n^2 q - \frac{11q^2}{4\sqrt{3}} + \frac{\sqrt{3}q^2}{4} + 2\sqrt{3}n^2 q A_\sigma - 2\sqrt{3}q^2 A_\sigma$$

$$-\frac{3}{2}\sqrt{3}q^2 A_\sigma^2 + \frac{n^2 \rho k_{NG}}{2\sqrt{3}} + \frac{1}{2}\sqrt{3}n^2\rho k_{NG} - \frac{3}{4}q\rho k_{NG} - \frac{11q\rho k_{NG}}{4\sqrt{3}} - \frac{1}{2}\sqrt{3}q\rho k_{NG}$$

$$-\frac{15}{8}q\rho A_\sigma k_{NG} - \frac{23}{8}\sqrt{3}q\rho A_\sigma k_{NG} - \frac{n^2 k_{PR}}{2} + \frac{n^2 k_{PR}}{2\sqrt{3}} - \frac{qk_{PR}}{4} - \frac{11qk_{PR}}{4\sqrt{3}} - \frac{9}{8}qA_\sigma k_{PR}$$

$$-\frac{17}{8}\sqrt{3}qA_\sigma k_{PR} - \frac{n^2\rho k_{NG}\Omega_g}{\sqrt{3}} - \sqrt{3}n^2\rho k_{NG}\Omega_g + \frac{3}{2}q\rho k_{NG}\Omega_g + \frac{11q\rho k_{NG}\Omega_g}{2\sqrt{3}}$$

$$+\sqrt{3}q\rho k_{NG}\Omega_g + \frac{15}{4}q\rho A_\sigma k_{NG}\Omega_g + \frac{23}{4}\sqrt{3}q\rho A_\sigma k_{NG}\Omega_g + \frac{n^2\rho k_{NG}\Omega_g^2}{2\sqrt{3}}$$

$$+\frac{1}{2}\sqrt{3}n^2\rho k_{NG}\Omega_g^2 - \frac{3}{4}q\rho k_{NG}\Omega_g^2 - \frac{11q\rho k_{NG}\Omega_g^2}{4\sqrt{3}} - \frac{1}{2}\sqrt{3}q\rho k_{NG}\Omega_g^2$$

$$-\frac{15}{8}q\rho A_\sigma k_{NG}\Omega_g^2 - \frac{23}{8}\sqrt{3}q\rho A_\sigma k_{NG}\Omega_g^2$$

$$\mathcal{D}_{2,1} = \sqrt{3}n^2 + \frac{n^4}{\sqrt{3}} + \frac{1}{2}\sqrt{3}q - \frac{35n^2q}{4\sqrt{3}} - \frac{1}{4}\sqrt{3}n^2q + \frac{11q^2}{2\sqrt{3}} - \frac{1}{2}\sqrt{3}q^2 - \frac{1}{2}\sqrt{3}n^2 A_\gamma$$

$$+ \frac{1}{2}\sqrt{3}qA_\gamma + \frac{9}{4}\sqrt{3}qA_\sigma - \frac{25}{4}\sqrt{3}n^2qA_\sigma + 4\sqrt{3}q^2 A_\sigma + \frac{3}{4}\sqrt{3}qA_\gamma A_\sigma + 3\sqrt{3}q^2 A_\sigma^2$$

$$+ \rho k_{NG} - \frac{\rho k_{NG}}{2\sqrt{3}} - \frac{1}{2}\sqrt{3}\rho k_{NG} - \frac{1}{4}n^2\rho k_{NG} - \frac{5}{4}\sqrt{3}n^2 \rho k_{NG} + \frac{9}{4}q\rho k_{NG}$$

$$+ \frac{11q\rho k_{NG}}{4\sqrt{3}} + 4\sqrt{3}q\rho k_{NG} + \frac{27}{8}\rho A_\gamma k_{NG} + \frac{23}{8}\sqrt{3}\rho A_\gamma k_{NG} + \frac{51}{8}q\rho A_\sigma k_{NG}$$

$$+ \frac{83}{8}\sqrt{3}q\rho A_\sigma k_{NG} + \frac{k_{PR}}{2} - \frac{k_{PR}}{2\sqrt{3}} + 2n^2 k_{PR} - \frac{n^2 k_{PR}}{\sqrt{3}} + \sqrt{3}n^2 k_{PR} - \frac{1}{4}q k_{PR}$$

$$+ \frac{7}{4}\sqrt{3}q k_{PR} + \frac{21}{8}A_\gamma k_{PR} + \frac{17}{8}\sqrt{3}A_\gamma k_{PR} + \frac{15}{8}qA_\sigma k_{PR} + \frac{39}{8}\sqrt{3}qA_\sigma k_{PR}$$

$$- 2\rho k_{NG}\Omega_g + \frac{\rho k_{NG}\Omega_g}{\sqrt{3}} + \sqrt{3}\rho k_{NG}\Omega_g + \frac{1}{2}n^2\rho k_{NG}\Omega_g + \frac{5}{2}\sqrt{3}n^2\rho k_{NG}\Omega_g$$

$$- \frac{9}{2}q\rho k_{NG}\Omega_g - \frac{11q\rho k_{NG}\Omega_g}{2\sqrt{3}} - 8\sqrt{3}q\rho k_{NG}\Omega_g - \frac{27}{4}\rho A_\gamma k_{NG}\Omega_g$$

$$- \frac{23}{4}\sqrt{3}\rho A_\gamma k_{NG}\Omega_g - \frac{51}{4}q\rho A_\sigma k_{NG}\Omega_g - \frac{83}{4}\sqrt{3}q\rho A_\sigma k_{NG}\Omega_g + \rho k_{NG}\Omega_g^2$$

$$- \frac{\rho k_{NG}\Omega_g^2}{2\sqrt{3}} - \frac{1}{2}\sqrt{3}\rho k_{NG}\Omega_g^2 - \frac{1}{4}n^2\rho k_{NG}\Omega_g^2 - \frac{5}{4}\sqrt{3}n^2\rho k_{NG}\Omega_g^2 + \frac{9}{4}q\rho k_{NG}\Omega_g^2$$

$$+ \frac{11q\rho k_{NG}\Omega_g^2}{4\sqrt{3}} + 4\sqrt{3}q\rho k_{NG}\Omega_g^2 + \frac{27}{8}\rho A_\gamma k_{NG}\Omega_g^2 + \frac{23}{8}\sqrt{3}\rho A_\gamma k_{NG}\Omega_g^2$$

$$+ \frac{51}{8}q\rho A_\sigma k_{NG}\Omega_g^2 + \frac{83}{8}\sqrt{3}q\rho A_\sigma k_{NG}\Omega_g^2$$

$$\mathcal{D}_{2,2} = \frac{1}{2\sqrt{3}} - \frac{\sqrt{3}}{2} - \frac{n^2}{\sqrt{3}} - \frac{1}{2}\sqrt{3}q + \frac{11n^2q}{2\sqrt{3}} - \frac{11q^2}{4\sqrt{3}} + \frac{1}{4}\sqrt{3}q^2 - \frac{15}{4}\sqrt{3}A_\gamma$$

$$+ \frac{17}{4}\sqrt{3}n^2 A_\gamma - \frac{1}{2}\sqrt{3}qA_\gamma + \frac{3}{4}\sqrt{3}A_\gamma^2 - \frac{9}{4}\sqrt{3}qA_\sigma + \frac{17}{4}\sqrt{3}n^2qA_\sigma - 2\sqrt{3}q^2 A_\sigma$$

$$- \frac{3}{4}\sqrt{3}qA_\gamma A_\sigma - \frac{3}{2}\sqrt{3}q^2 A_\sigma^2 - \frac{9}{4}\rho k_{NG} + \frac{5}{4}\sqrt{3}\rho k_{NG} + n^2\rho k_{NG} + \frac{3}{2}\sqrt{3}n^2\rho k_{NG}$$

$$- \frac{13}{4}q\rho k_{NG} - \frac{29}{4}\sqrt{3}q\rho k_{NG} - 9\rho A_\gamma k_{NG} - \frac{15}{2}\sqrt{3}\rho A_\gamma k_{NG} - \frac{75}{8}q\rho A_\sigma k_{NG}$$

$$-\frac{123}{8}\sqrt{3}q\rho A_\sigma k_{NG} + \frac{k_{PR}}{\sqrt{3}} - \sqrt{3}k_{PR} - 2n^2 k_{PR} + \frac{1}{2}qk_{PR} - \frac{11qk_{PR}}{2\sqrt{3}}$$

$$+\sqrt{3}qk_{PR} - \frac{15}{4}A_\gamma k_{PR} - \frac{11}{4}\sqrt{3}A_\gamma k_{PR} - \frac{3}{4}qA_\sigma k_{PR} - \frac{11}{4}\sqrt{3}qA_\sigma k_{PR}$$

$$+\frac{9}{2}\rho k_{NG}\Omega_g - \frac{5}{2}\sqrt{3}\rho k_{NG}\Omega_g - 2n^2\rho k_{NG}\Omega_g - 3\sqrt{3}n^2\rho k_{NG}\Omega_g$$

$$+\frac{13}{2}q\rho k_{NG}\Omega_g + \frac{29}{2}\sqrt{3}q\rho k_{NG}\Omega_g + 18\rho A_\gamma k_{NG}\Omega_g + 15\sqrt{3}\rho A_\gamma k_{NG}\Omega_g$$

$$+\frac{75}{4}q\rho A_\sigma k_{NG}\Omega_g + \frac{123}{4}\sqrt{3}q\rho A_\sigma k_{NG}\Omega_g - \frac{9}{4}\rho k_{NG}\Omega_g^2 + \frac{5}{4}\sqrt{3}\rho k_{NG}\Omega_g^2$$

$$+n^2\rho k_{NG}\Omega_g^2 + \frac{3}{2}\sqrt{3}n^2\rho k_{NG}\Omega_g^2 - \frac{13}{4}q\rho k_{NG}\Omega_g^2 - \frac{29}{4}\sqrt{3}q\rho k_{NG}\Omega_g^2$$

$$-9\rho A_\gamma k_{NG}\Omega_g^2 - \frac{15}{2}\sqrt{3}\rho A_\gamma k_{NG}\Omega_g^2 - \frac{75}{8}q\rho A_\sigma k_{NG}\Omega_g^2 - \frac{123}{8}\sqrt{3}q\rho A_\sigma k_{NG}\Omega_g^2$$

$$\mathcal{D}_{2,3} = \frac{5}{2}\rho k_{NG} - \frac{3}{2}\sqrt{3}\rho k_{NG} - n^2\rho k_{NG} - \frac{n^2\rho k_{NG}}{\sqrt{3}} + \frac{9}{4}q\rho k_{NG} + \frac{11q\rho k_{NG}}{2\sqrt{3}}$$

$$+\frac{15}{4}\sqrt{3}q\rho k_{NG} + \frac{75}{8}\rho A_\gamma k_{NG} + \frac{63}{8}\sqrt{3}\rho A_\gamma k_{NG} + \frac{57}{8}q\rho A_\sigma k_{NG} + \frac{97}{8}\sqrt{3}q\rho A_\sigma k_{NG}$$

$$-5\rho k_{NG}\Omega_g + 3\sqrt{3}\rho k_{NG}\Omega_g + 2n^2\rho k_{NG}\Omega_g + \frac{2n^2\rho k_{NG}\Omega_g}{\sqrt{3}} - \frac{9}{2}q\rho k_{NG}\Omega_g$$

$$-\frac{11q\rho k_{NG}\Omega_g}{\sqrt{3}} - \frac{15}{2}\sqrt{3}q\rho k_{NG}\Omega_g - \frac{75}{4}\rho A_\gamma k_{NG}\Omega_g - \frac{63}{4}\sqrt{3}\rho A_\gamma k_{NG}\Omega_g$$

$$-\frac{57}{4}q\rho A_\sigma k_{NG}\Omega_g - \frac{97}{4}\sqrt{3}q\rho A_\sigma k_{NG}\Omega_g + \frac{5}{2}\rho k_{NG}\Omega_g^2 - \frac{3}{2}\sqrt{3}\rho k_{NG}\Omega_g^2$$

$$-n^2\rho k_{NG}\Omega_g^2 - \frac{n^2\rho k_{NG}\Omega_g^2}{\sqrt{3}} + \frac{9}{4}q\rho k_{NG}\Omega_g^2 + \frac{11q\rho k_{NG}\Omega_g^2}{2\sqrt{3}} + \frac{15}{4}\sqrt{3}q\rho k_{NG}\Omega_g^2$$

$$+\frac{75}{8}\rho A_\gamma k_{NG}\Omega_g^2 + \frac{63}{8}\sqrt{3}\rho A_\gamma k_{NG}\Omega_g^2 + \frac{57}{8}q\rho A_\sigma k_{NG}\Omega_g^2 + \frac{97}{8}\sqrt{3}q\rho A_\sigma k_{NG}\Omega_g^2$$

$$\mathcal{D}_{2,4} = -\rho k_{NG} + \frac{\rho k_{NG}}{\sqrt{3}} - \frac{1}{2}q\rho k_{NG} - \frac{11q\rho k_{NG}}{2\sqrt{3}} - \frac{21}{4}\rho A_\gamma k_{NG} - \frac{17}{4}\sqrt{3}\rho A_\gamma k_{NG}$$

$$-\frac{9}{4}q\rho A_\sigma k_{NG} - \frac{17}{4}\sqrt{3}q\rho A_\sigma k_{NG} + 2\rho k_{NG}\Omega_g - \frac{2\rho k_{NG}\Omega_g}{\sqrt{3}} + q\rho k_{NG}\Omega_g$$

$$+ + \frac{11q\rho k_{NG}\Omega_g}{\sqrt{3}} \frac{21}{2}\rho A_\gamma k_{NG}\Omega_g + \frac{17}{2}\sqrt{3}\rho A_\gamma k_{NG}\Omega_g$$

$$+ \frac{9}{2}q\rho A_\sigma k_{NG}\Omega_g + \frac{17}{2}\sqrt{3}q\rho A_\sigma k_{NG}\Omega_g - \rho k_{NG}\Omega_g^2 + \frac{\rho k_{NG}\Omega_g^2}{\sqrt{3}}$$

$$- \frac{1}{2}q\rho k_{NG}\Omega_g^2 - \frac{11q\rho k_{NG}\Omega_g^2}{2\sqrt{3}} - \frac{21}{4}\rho A_\gamma k_{NG}\Omega_g^2 - \frac{17}{4}\sqrt{3}\rho A_\gamma k_{NG}\Omega_g^2$$

$$- \frac{9}{4}q\rho A_\sigma k_{NG}\Omega_g^2 - \frac{17}{4}\sqrt{3}q\rho A_\sigma k_{NG}\Omega_g^2$$